\documentclass{jfm}

\usepackage{hyperref,graphicx,pstool,amsmath,amsfonts,amssymb,booktabs,overpic,multirow,rotating}

\usepackage{tikz}
\usetikzlibrary{intersections,calc,shapes.arrows,shapes.geometric,plotmarks}

\shorttitle{Wave energy absorption by a floating air bag}
\shortauthor{A. Kurniawan, J. R. Chaplin, D. M. Greaves, and M. Hann}

\title{Wave energy absorption by a floating air bag}
\author{A. Kurniawan\aff{1,3}
  \corresp{\email{aku@civil.aau.dk}}, 
J. R. Chaplin\aff{2},
 D. M. Greaves\aff{1},
  \and    M. Hann\aff{1}}

\affiliation{\aff{1}School of Marine Science and Engineering, Plymouth University, Drake Circus, \\Plymouth, PL4 8AA, UK
\aff{2}Faculty of Engineering and the Environment, University of Southampton, Highfield,\\ Southampton, SO17 1BJ, UK
\aff{3}Department of Civil Engineering, Aalborg University, Thomas Manns Vej 23, \\9220 Aalborg, Denmark}

\begin{document}

\maketitle

\begin{abstract}
A floating air bag, ballasted in water, expands and contracts as it heaves under wave action. 
Connecting the bag to a secondary volume via a turbine transforms the bag into a device capable of generating useful energy from the waves. 
Small-scale measurements of the device reveal some interesting properties, which are successfully predicted numerically. 
Owing to its compressibility, the device can have a heave resonance period longer than that of a rigid device of the same shape and size, without any phase control. 
Furthermore, varying the amount of air in the bag is found to change its shape and hence its dynamic response, while varying the turbine damping or the air volume ratio changes the dynamic response without changing the shape. 
\end{abstract}


\section{Introduction}

Wave energy is still searching for an economic solution. 
Some efforts to make wave energy economic are targeted at engineering the dynamic response of a wave energy device so that it 
extracts more energy out of the waves relative to the investment put in.
Recent strategies include introducing multiple resonances  to broaden the device resonance bandwidth~\citep{EvansPorter2012, Crowley2013} and  incorporating a negative spring mechanism to lengthen the device resonance period while broadening its bandwidth at the same time~\citep{Zhang2014,Todalshaug2016}. 
This paper is concerned with another related approach, that is, using a flexible deformable body as a wave energy device so that it can have a longer natural period compared to that of a rigid body of the same size. 
A flexible body also gives the device the ability to change its mean shape to adapt to different environmental conditions.

One practical way of incorporating a deformable geometry into a wave energy device was proposed by~\citet{Farley2011,Farley2012patent}. 
The floating device resembles a pair of bellows that are hinged at the bottom, giving a V-shaped cross-section. 
As the device heaves under wave action, the bellows expand and contract, creating an air flow into and out of a separate volume through a turbine. 
The increase in heave resonance period depends on the restoring stiffness of the plates, which is provided by compression of the enclosed air.



\begin{figure}
\centering
\includegraphics[scale=1]{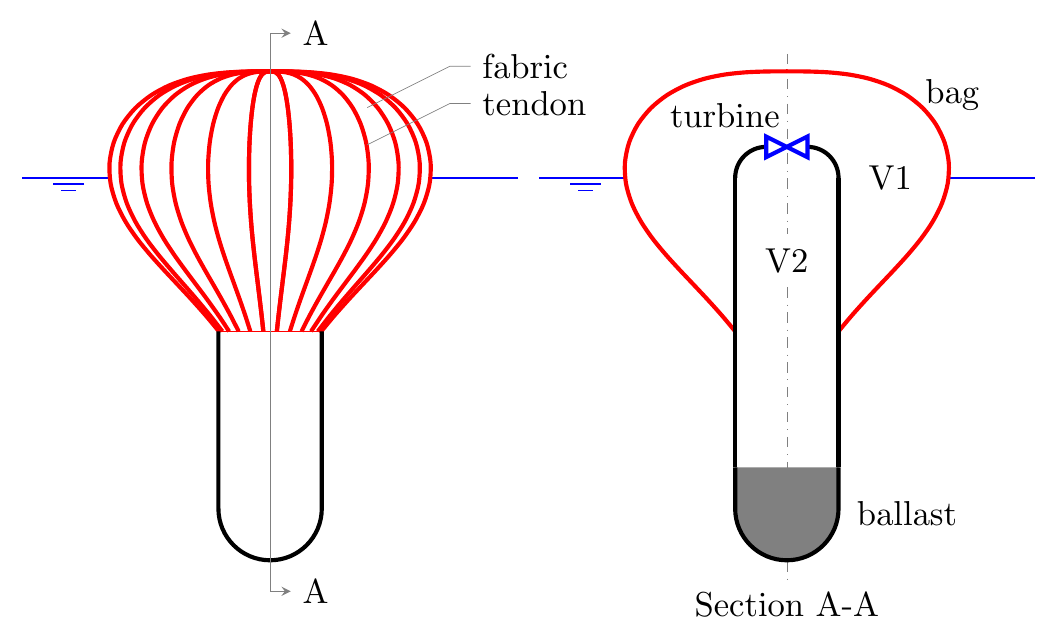}
\caption{Sketch of the device with the main parts identified.}
\label{sq1device}
\end{figure}

More recently, attention has turned to an axisymmetric version of the device (figure~\ref{sq1device}).
The pair of rigid plates is replaced by a completely flexible bag, thus eliminating the need for hinges. 
The bag exchanges air through a turbine with a secondary volume housed within the device, making it an entirely sealed system.
The construction of the bag is that of a fabric encased within an array of relatively stiff tendons. 
When inflated, the bag forms meridional lobes between the tendons, effectively keeping the tension in the fabric to a minimum, while the tendons carry most of the tension. 
Such bags have been used mainly for aerospace applications~\citep{Pagitz2007}, but recently also underwater~\citep{Pimm2014}.
When the internal-external pressure difference is uniform, such as in air, the bag takes a pumpkin-like shape~\citep{Taylor1919}, but when immersed in water, the shape is more like that of an inverted pear due to the increasing external pressure with depth.


It is of interest to predict the response of this device in waves, including its power absorption characteristics. 
The use of flexible fabric structures for wave energy extraction is actually not new and has been pioneered by~\citet{French1979} and~\citet{Bellamy1982}. 
However, a rigorous analysis of their static and dynamic behaviour, including their interaction with waves, is still lacking, and moreover, the construction and geometry of their bags are different from that used in the present device. 
To this end, a series of studies has been devoted to developing an understanding of the static and dynamic behaviour of the device. 
A numerical method for calculating the equilibrium shape of the bag in still water was first developed and then validated through model tests~\citep{Kurniawan2015b,Chaplin2015}.
Subsequently, a linear frequency-domain model was developed to predict the response of the device when forced to oscillate in otherwise still water, by means of a pump, in place of a turbine. 
The predictions agreed reasonably well with measurements~\citep{Chaplin2015}.
In~\citet{Kurniawan2016}, measurements of the device's response in regular waves were reported, but numerical predictions were presented only for the ratio of the pressure in the bag to that in the secondary volume, and the resonance period in heave.

This paper summarises all these previous studies and presents the analyses and results in a more complete manner. 
To understand the characteristics of such deformable body as a wave energy device, we first analyse an ideal semisubmerged sphere which deforms by uniform expansion and contraction while oscillating in heave. 
The deformation of the floating bag is not uniform, but the same general characteristics apply. 
A linear frequency-domain model for predicting the response of the floating bag device when absorbing energy in waves is presented for the first time,  extending the calculation followed previously for the pumped case.
The predictions are generally found to be in good agreement with new measurements in waves.
With confidence in the numerical model, a parametric study explores the device's behaviour over a range of conditions.
Some practical implications are suggested. 

\section{Heaving pulsating sphere}

We first consider a semisubmerged sphere of radius $a$ that expands and contracts uniformly as it heaves.
The pulsating mode $(i = 7)$ and the heave mode $(i = 3)$ are coupled through the following equations of motion:
\begin{equation}
\left[
-\omega^2 \begin{pmatrix}
M + A_{33} & A_{37} \\
A_{73} & A_{77}
\end{pmatrix} + \mathrm{i}\omega \begin{pmatrix}
B_{33} & B_{37} \\
B_{73} & B_{77}
\end{pmatrix} + \begin{pmatrix}
C_{33} & C_{37} \\
C_{73} & C_{77} + K_{77}
\end{pmatrix} 
\right] 
\begin{Bmatrix}
\xi_3 \\ \xi_7
\end{Bmatrix} = 
\begin{Bmatrix}
F^{\mathrm{exc}}_3 \\ F^{\mathrm{exc}}_7
\end{Bmatrix},
\end{equation}
where $\omega$ is the angular frequency, $M$ is the mass of the sphere, $A_{ij}$, $B_{ij}$, and $C_{ij}$ are the added mass, radiation damping, and hydrostatic stiffness in mode $i$ due to motion in mode $j$.  
Furthermore, $F^{\mathrm{exc}}_i$ and $\xi_i$ are the wave excitation force and the displacement in mode $i$, and $K_{77}$ is the internal stiffness of the pulsation.

If the pulsating mode is stiffer than the heaving mode, i.e. $C_{77} + K_{77} > C_{33}$, then expansion is in phase with upward motion, or $\xi_7 = r \, \xi_3$, where $r$ is real and positive, and is a measure of the pulsation compliance of the sphere.
The heave natural frequency of the pulsating sphere is therefore given as 
\begin{equation}
\omega_{03}^2 = \frac{C_{33} + C_{37} \, r}{M + A_{33} + A_{37} \, r} . \label{natfreqpuls}
\end{equation}

Following~\citet{Newman1994}, the hydrostatic stiffness $C_{ij}$ can be expressed as an integral over the mean wetted surface of the body $S_B$:
\begin{equation}
C_{ij} = \rho g \int_{S_B} n_j (w_i + z D_i) \mathrm{d}S , \label{Cij}
\end{equation}
where $\rho$ is the water density, $g$ is the acceleration due to gravity, and
$n_j$, $w_j$, and $D_j$ are the normal component, the vertical component, and the divergence, of mode $j$.
The pulsating mode of the sphere can be written in Cartesian components as $(x \boldsymbol{i} + y \boldsymbol{j} + z \boldsymbol{k})/a$.
Thus, from~\eqref{Cij} we obtain $C_{33} = \rho g \pi a^2$ and $C_{37} = -2 \rho g \pi a^2$ (this is also obvious from geometry).
Substituting these into~\eqref{natfreqpuls}, we find that the sphere is stable in heave if $r < 0.5$.

\begin{figure}
\centering
\includegraphics[trim = 0mm 3mm 0mm 2mm, clip, scale=1]{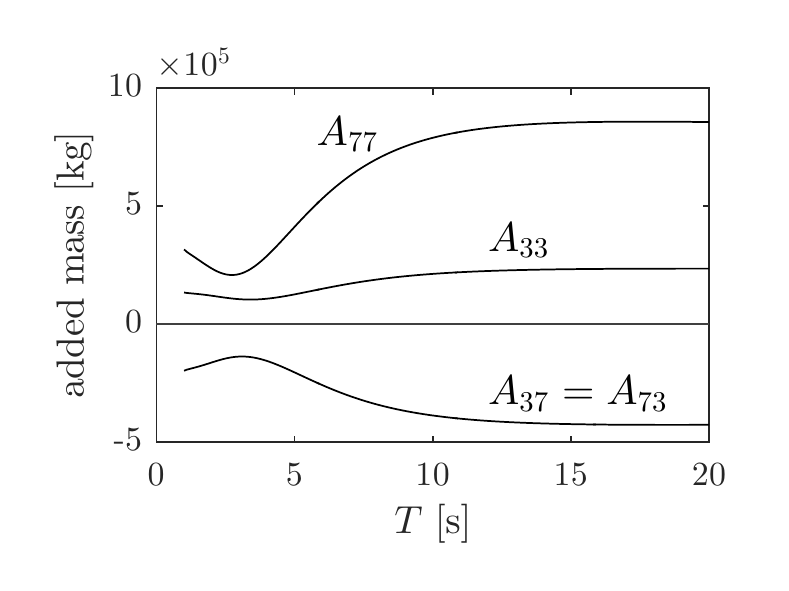}
\caption{Added mass of a pulsating semisubmerged sphere with radius 5 m, calculated using~\citet{WAMIT7.1}. }
\label{addedm_puls_sphere}
\end{figure}

\begin{figure}
\centering
\includegraphics[trim = 0mm 3mm 0mm 2mm, clip, scale=1]{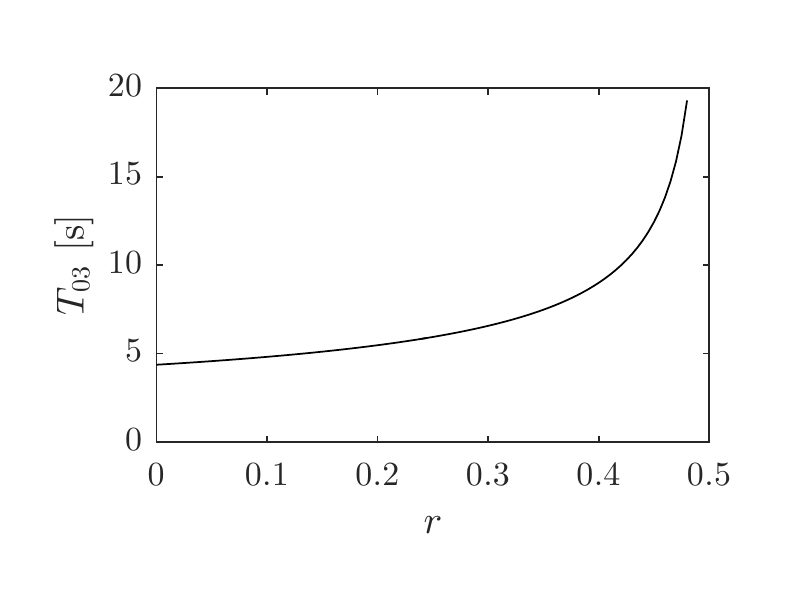}
\caption{Variation of heave natural period with $r$, for a pulsating semisubmerged sphere with radius 5 m. }
\label{natper_puls_sphere}
\end{figure}

The added masses $A_{33}$ and $A_{37}$ can be evaluated using various methods~\citep[e.g.][]{Hulme1982,Lopes2002}. 
As seen from figure~\ref{addedm_puls_sphere}, the added mass $A_{37}$ is negative, but $A_{33} + A_{37} \, r$ is positive for $r < 0.5$. 
Therefore, from~\eqref{natfreqpuls} we have $0 < \omega_{03} \leq \sqrt{\rho g \pi a^2 / (M + A_{33})}$.
The upper limit is just the heave natural frequency of a rigid sphere.  
The variation of the heave natural period of the pulsating sphere with $r$, for a sphere of radius $a = 5$ m, is plotted in figure~\ref{natper_puls_sphere}.

From geometry, $F^{\mathrm{exc}}_7 = -2 F^{\mathrm{exc}}_3$. 
The negative sign is due to the two forces being in antiphase, and the factor of 2  to the wetted surface area of the semisubmerged sphere being twice its water plane area.
By a well-known reciprocity relation between wave excitation force and radiation damping~\citep{Newman1962}, the radiation damping $B_{37} = B_{73}$ is therefore negative. 
It follows that the total energy radiated by the heaving pulsating sphere is less than that radiated by a heaving rigid sphere of the same dimension, for the same amplitude of the heave motion. 
If the pulsating sphere were to act as a wave energy absorber, then we expect that while it can resonate at a longer period, its resonance bandwidth would be less than that of a rigid sphere of the same size. 
However, the relative bandwidth of the absorbed power is 
related to the ratio of the damping of the system to the square root of the product between the stiffness and mass of the system~\citep[see][]{Falnes2002}.
Since not only the radiation damping is less but also the total stiffness and mass of the pulsating sphere are less (due to the negative off-diagonal terms) than those of the rigid sphere, the reduction in bandwidth will not be severe.   

Having analysed the pulsating sphere, we will now turn our attention to the floating air bag, where we will show that the same general characteristics apply.

\section{Theory}

\subsection{Static shape calculations} \label{static_calc}

The static shape of the bag in still water is determined uniquely by (for example) the submergence of the bottom of the bag and the static internal pressure. 
To calculate the shape of the bag, we assume that all tension is carried by the tendons while the tension in the fabric is zero.
This assumption is equivalent to having infinitely many tendons, and when the internal-external pressure difference is uniform, the shape of the bag is that of an isotensoid~\citep{Taylor1919}.
Since the bag is axisymmetric, it is sufficient to calculate the profile of a single tendon. 
A method of calculation has been described by~\citet{Kurniawan2015b} and is presented here for completeness.
In sum, the process involves discretising the tendon into  $N$ arc elements with known lengths but unknown radii of curvature. 
The radius of curvature of each element is obtained by satisfying the force equilibrium normal to the element, as we progress element by element from the top of the bag to the bottom.
As the tendon tension and the top elevation are not known beforehand, an iteration is necessary in order to obtain the correct tension and elevation.
The problem is akin to that of inflatable dams under static loading, and the solution procedure is similar to those used by~\citet{Harrison1970} and~\citet{Parbery1976}, except that in the present problem we add further simplifying assumptions that the tendons are massless and inextensible. 

\begin{figure}
\centering
\includegraphics[trim = 0mm 5mm 0mm 0mm, clip, scale=0.9]{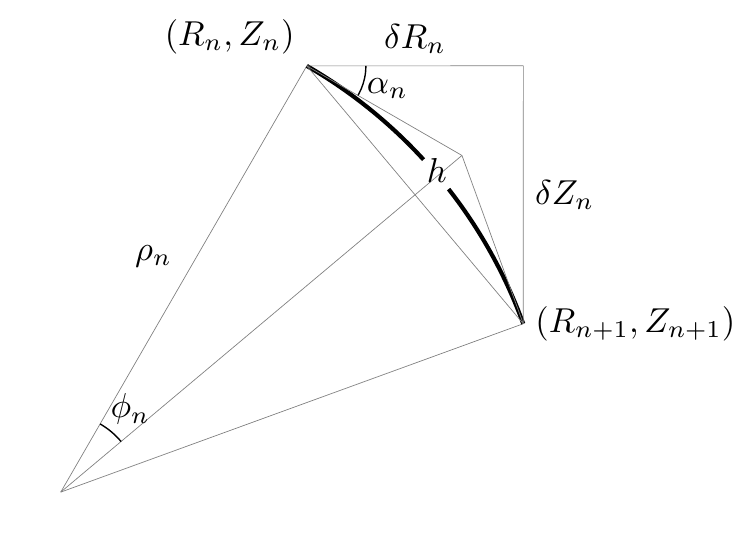}
\caption{One arc element. }
\label{sketch_arc_element}
\end{figure}

We define a cylindrical coordinate system $(R, \theta, Z)$ where $Z=0$ is the mean free surface and the $Z$-axis is pointing up and coincident with the device's vertical axis.
For simplicity, all $N$ elements have uniform length $h$.
Looking at figure~\ref{sketch_arc_element}, we see that the arc length $h$ is related to its radius of curvature $\rho_n$ through 
\begin{equation}
h = -2\rho_n \phi_n, \label{h-rho}
\end{equation}
where $2\phi_n = \delta \alpha_n$ is the arc sector angle.
The radius of curvature $\rho_n$ is defined to be positive when the fabric is bulging outwards.
The arc chord length is  $(h / \phi_n) \sin \phi_n$. 
Thus, 
\begin{gather}
\delta R_n = (h / \phi_n) \sin\phi_n \cos (\alpha_n + \phi_n) \label{dr-b}\\
\delta Z_n = (h / \phi_n) \sin\phi_n \sin(\alpha_n + \phi_n) . \label{dz-b}
\end{gather}
For small $\phi_n$, the chord length approximates to the arc length, and the above equations reduce to
\begin{gather}
\delta R_n \approx h (\cos \alpha_n - \phi_n \sin \alpha_n) \label{dr}\\
\delta Z_n \approx h (\sin \alpha_n + \phi_n \cos \alpha_n) . \label{dz}
\end{gather}

For a given internal pressure $P$ (excluding atmospheric) and submergence of the bottom of the bag $Z_{\text{bot}}$, 
we start by making guesses of the sum of tension in all tendons $T$, and the elevation of the top of the bag $Z_1$.
The calculation proceeds from here, where $R_1 = R_\mathrm{top}$ and $\alpha_1 = 0$.
As a first approximation, $\phi_1 \approx 0$, thus $\delta R_1 \approx h$ and $\delta Z_1 \approx 0$ according to~\eqref{dr} and~\eqref{dz}.
The radius of curvature $\rho_1$ is calculated from the balance of normal forces; in the general case,
\begin{equation}
\rho_n = \frac{T}{2 \pi (P + H_{n+0.5} \rho g Z_{n+0.5}) R_{n+0.5}} , \label{rho}
\end{equation}
where $R_{n+0.5}$ and $Z_{n+0.5}$ are the radius and elevation at midpoint of element $n$, and 
\begin{equation}
H_{n+0.5} = \begin{cases}
1, &  Z_{n+0.5} < 0 \\
0, &  Z_{n+0.5} \geq 0 .
\end{cases}
\end{equation}  
For the $n$th element,
knowing the radius of curvature $\rho_n$, we may evaluate the angle $\phi_n$ from~\eqref{h-rho} and have better approximations for $\delta R_n$ and $\delta Z_n$ from~\eqref{dr-b} and~\eqref{dz-b}.
These, in turn, give better approximations for $R_{n+0.5}$, $Z_{n+0.5}$, $\rho_n$, and $\phi_n$, which again are used in~\eqref{dr-b} and~\eqref{dz-b} to give the final $\delta R_n$ and $\delta Z_n$ values needed to proceed to the next node along the tendon towards the bottom.
A simple iterative procedure adjusts the starting values of $T$ and $Z_1$ to bring the last node $(R_{N+1},Z_{N+1})$ to the specified $(R_{\text{bot}},Z_{\text{bot}})$ at the bottom of the bag.


Denoting $B$ as the buoyancy of the bag, and neglecting its weight and that of the enclosed air, we have 
\begin{equation}
B + T \sin \alpha_{N+1} + \pi R_{N+1}^2 (P + \rho g Z_{N+1}) = 0  \label{displacement_check}
\end{equation}
from static equilibrium at the bottom of the bag. 
This equation provides a check on the calculated shape of the bag as well as the final value of tension $T$. 


\begin{figure}
\centering
\begin{overpic}[trim = 0mm 5mm 0mm 6mm, clip]{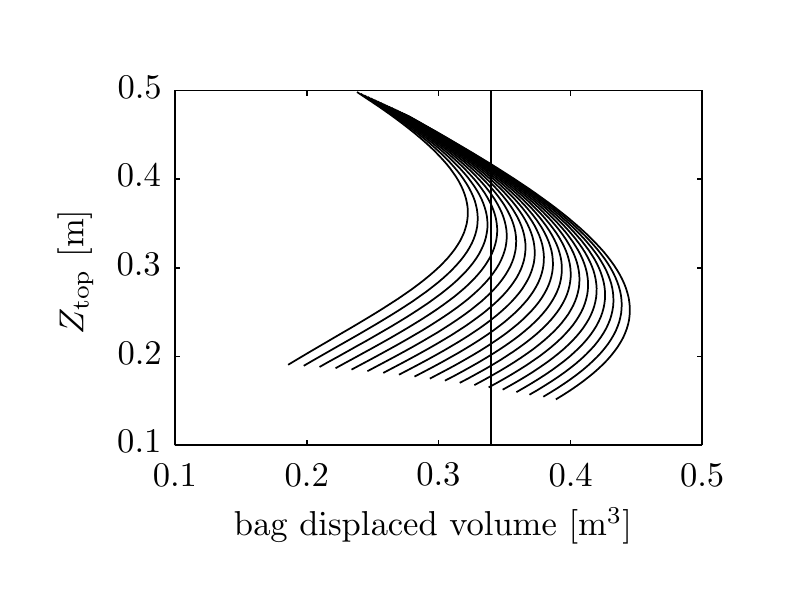} 
\put(25,14){\begin{tikzpicture}[auto,scale=.5]
\draw [->] (0,0) .. controls (1,0) and (2,0) .. (5,-0.4);
\draw (2.5,-.8) node {\small increasing pressure};
\end{tikzpicture}}
\end{overpic}
\caption{Typical plots of top elevation vs. displaced volume of the bag. 
Intersections of the curves with a vertical line give the possible top elevations of the bag for the same ballast, for different amounts of air in the bag.
This plot refers to the condition of the larger bag used in experiments discussed later.}
\label{elevation-displacedvolume}
\end{figure}
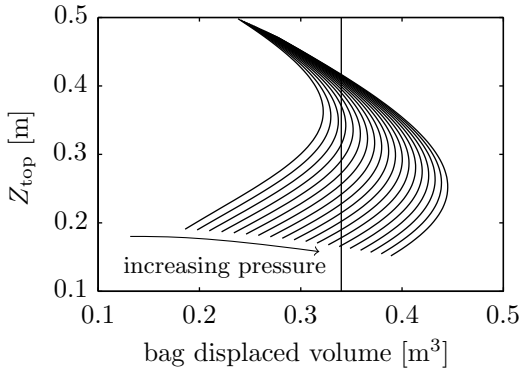
 
Now, the above calculations give the shape of the bag when the bottom elevation and internal pressure are specified. 
It is found that for a given internal pressure there are in general two solutions that provide the same displaced volume.
This is illustrated by figure~\ref{elevation-displacedvolume}, where each curve shows the variation of the top elevation and the displaced volume of the bag as its bottom elevation is varied while the internal pressure is kept constant. 
Since for a free-floating device the buoyancy of the bag must equal the submerged weight of the ballast, intersections of these curves with a vertical line give the possible top elevations of the bag (and the corresponding shapes) for a given ballast.
For a certain range of pressures, a floating bag with a given ballast can have two different equilibrium shapes for the same pressure. 
The two shapes have the same displaced volume as set by the ballast, but the total volumes of the bag and hence the air masses are different. 
More of this will be discussed in section~\ref{static_results}.

\subsection{Linear frequency-domain model}


\subsubsection{Governing equations for the bag and ballast}

\begin{figure}
\centering
\includegraphics[trim = 0mm 0mm 0mm 0mm, clip, scale=0.9]{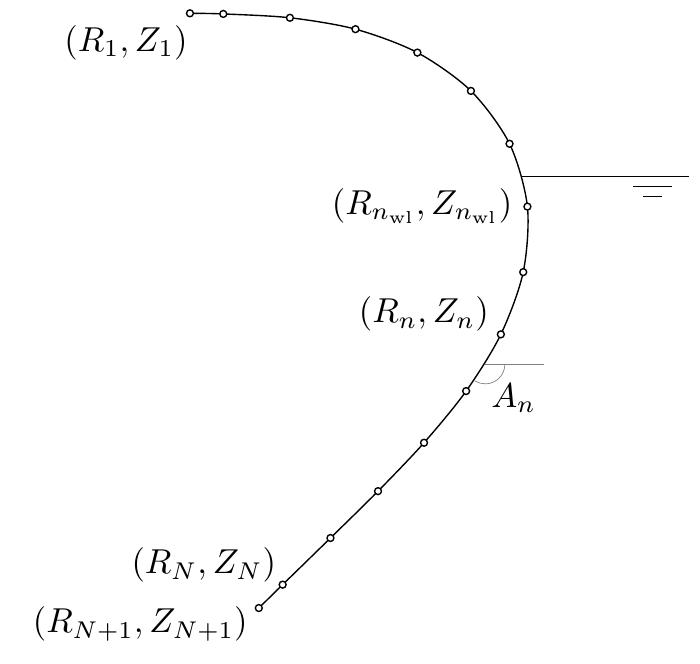} 
\caption{Discretised tendon.}
\label{discretised_tendon}
\end{figure}

Having calculated the static shape with equally spaced nodal coordinates $(R_n,Z_n), n=1,\dots,N+1$, it is helpful to define a new set of $N'+1 = N+2$ nodes where 
\begin{equation}
(R'_n,Z'_n) = \begin{cases}
(R_1,Z_1), &  n = 1 \\
(R_{n-0.5},Z_{n-0.5}), &  n = 2,\dots,N+1 \\
(R_{N+1},Z_{N+1}), &  n = N+2
\end{cases},
\end{equation}
and to define a set of element angles $A_n$, where $A_n = \alpha_n, n=1,\dots,N'$.
The new distribution of nodes is then like that in figure~\ref{discretised_tendon}, where we have omitted the prime symbols to simplify notations and will continue to do so in the following.

At static equilibrium, the sum of forces normal to each node of the tendon is zero:
\begin{equation}
2\pi h (P + H_n \rho g Z_n) R_n = T (A_{n-1} - A_n) . \label{tendon_eq_static}
\end{equation}
This equation is equivalent to~\eqref{rho}, where~\eqref{h-rho} has been used to express the radius of curvature in terms of the element angles. 
As before, $h$ is the element length, $P$ is the static bag pressure, and $T$ is the sum of static tension in all tendons.

On the ballast, static equilibrium requires that
\begin{equation}
W_\mathrm{B} + T \sin A_{N} + \pi R_{N+1}^2 (P + \rho g Z_{N+1}) = 0 , \label{ballast_eq_static}
\end{equation}
where $W_\mathrm{B}$ is the submerged weight of the ballast. 

It is assumed that all excitations and the resulting displacements are harmonic and of small amplitude, allowing us to model the device in the frequency domain and to neglect terms of second and higher orders.
The bag and the ballast are assumed to oscillate around the mean, or static, position, and the time-dependent part of any oscillatory quantity can be expressed as the real part of the product of a complex amplitude and $\mathrm{e}^{\mathrm{i}\omega t}$.
Furthermore, only vertical and radial displacements are considered. 

Neglecting the inertia of each node, we thus have
\begin{equation} 
2\pi h \left[ P + p_1 + H_n \rho g (Z_n + z_n + \xi_3) - p_n^\mathrm{h} \right] (R_n + r_n) = (T + \tau) (A_{n-1} - A_n + a_{n-1} - a_n)  \label{tendon_eq_full}
\end{equation}
on each node of the tendon. 
The complex amplitudes of the time-dependent parts have been denoted by lower-case letters to distinguish them from the mean, or static parts, which are denoted by upper-case letters. 
Thus, $p_1$ is the pressure change in the bag; $\xi_3$ is the vertical displacement of the ballast; $r_n$ is the radial displacement of node $n$; $z_n$ is the vertical displacement of node $n$ relative to $\xi_3$; $a_n$ is the angular displacement of element $n$; $\tau$ is the change in the total tendon tension; and $p_n^\mathrm{h}$ is the hydrodynamic pressure on node $n$. 

On the ballast, we have 
\begin{equation}
-W_\mathrm{B} - (T + \tau) \sin (A_{N} + a_N) - \pi R_{N+1}^2 \left[P + p_1 + \rho g (Z_{N+1} + \xi_3)\right] + f_\mathrm{B}^\mathrm{h} = -\omega^2 M_\mathrm{B} \xi_3 ,  \label{ballast_eq_full}
\end{equation}
where  $f_\mathrm{B}^\mathrm{h}$ is the vertical hydrodynamic force on the ballast and $M_\mathrm{B}$ is the ballast mass. 

In the following derivations, only terms linear in the complex amplitudes $p_1$, $p_n^\mathrm{h}$, $f_\mathrm{B}^\mathrm{h}$, $r_n$, $z_n$, $\xi_3$, $\tau$, and $a_n$ are kept and higher-order terms are discarded. 

Subtracting~\eqref{tendon_eq_static} from~\eqref{tendon_eq_full} and keeping only first-order terms, we thus have
\begin{equation} 
2\pi h \left\{ ( P + H_n \rho g Z_n ) r_n + \left[ p_1 + H_n \rho g (z_n + \xi_3) - p_n^\mathrm{h} \right] R_n \right\} = T (a_{n-1} - a_n) + \tau (A_{n-1} - A_n)   \label{tendon_eq_dynamic}
\end{equation}
for each node of the tendon.
Likewise, subtracting~\eqref{ballast_eq_static} from~\eqref{ballast_eq_full}, we have
\begin{equation} 
-T a_N \cos A_N - \tau \sin A_N  - \pi R_{N+1}^2 (p_1 + \rho g \xi_3) + f_\mathrm{B}^\mathrm{h} = -\omega^2 M_\mathrm{B} \xi_3 .  \label{ballast_eq_dynamic}
\end{equation}
for the ballast.

Each of the upper-case quantities in~\eqref{tendon_eq_dynamic} and~\eqref{ballast_eq_dynamic} is either specified or a solution of the static calculation described in section~\ref{static_calc}, whereas
the complex amplitudes $p_1$, $p_n^\mathrm{h}$, $f_\mathrm{B}^\mathrm{h}$, $r_n$, $z_n$, $\xi_3$, $\tau$, and $a_n$ are as yet unknown.
However, these complex amplitudes are not independent and are related to one another.
These relationships, as described below, as well as the boundary conditions $r_1 = a_1 = 0$ at the top of the bag and $r_{N+1} = z_{N+1} = 0$ at the bottom,
render the number of available equations equal to the number of unknowns.

\subsubsection{Relationship between radial and vertical nodal displacements} \label{radial-vertical_rel}

\begin{figure}
\centering
\includegraphics[trim = 0mm 0mm 0mm 0mm, clip, scale=0.9]{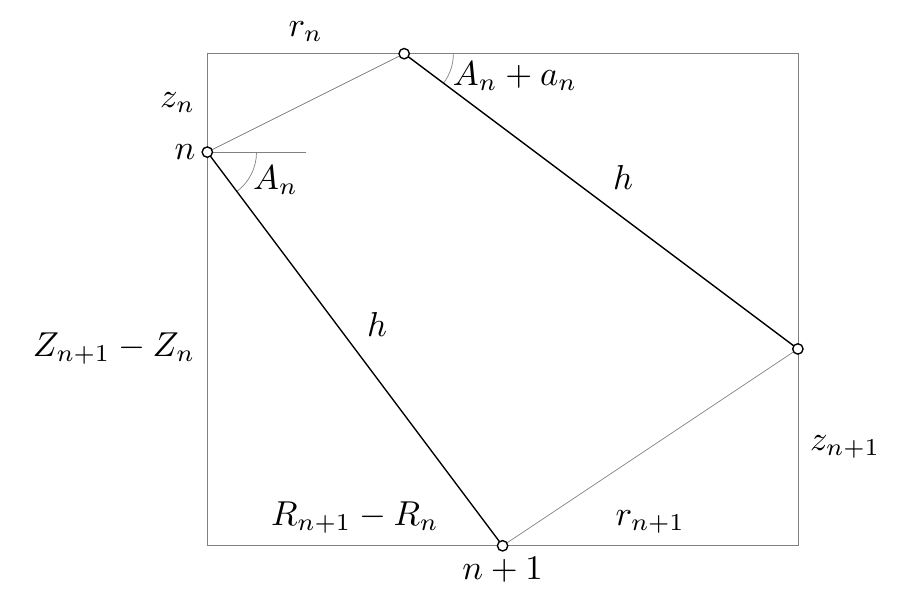} 
\caption{The distance between any two neighbouring nodes is assumed to be constant to first order.}
\label{tendon_element}
\end{figure}

The tendons are assumed to be inextensible, which means that the distance between any two neighbouring nodes does not change, assuming that the arc sector angles $\phi_n$ are so small that the arc chord length approximates to the arc length $h$.  
This provides a relationship between the radial and vertical displacements of the nodes.
Thus considering figure~\ref{tendon_element}, we have
\begin{equation}
(R_{n+1} + r_{n+1} - R_n - r_n)^2 + (Z_{n+1} + z_{n+1} - Z_n - z_n)^2 = (R_{n+1} - R_n)^2 + (Z_{n+1} - Z_n)^2 ,
\end{equation}
which, after discarding higher-order terms, reduces to 
\begin{equation}
z_{n+1} - z_n = -\frac{R_{n+1} - R_n}{Z_{n+1} - Z_n} (r_{n+1} - r_n) . \label{rz_relation}
\end{equation}

\subsubsection{Relationship between angular and radial displacements}

Likewise, the angular displacement $a_n$ can be expressed in terms of the radial displacements of the nodes.
According to figure~\ref{tendon_element}, we have
\begin{equation}
h \cos A_n = R_{n+1} - R_n  \label{element_static}
\end{equation}
at the original mean position of the element, and 
\begin{equation}
h \cos (A_n + a_n) = R_{n+1} - R_n + r_{n+1} - r_n  \label{element_full}
\end{equation}
at the displaced position.
Subtracting~\eqref{element_static} from~\eqref{element_full} and again discarding nonlinear terms yield
\begin{equation}
a_n = -\frac{r_{n+1} - r_n}{h \sin A_n} = -\frac{r_{n+1} - r_n}{Z_{n+1} - Z_n} . \label{ar_relation}
\end{equation}

\subsubsection{Relationship between pressure change and nodal displacements}

The pressure amplitude $p_1$ is related to the volume amplitude of the bag $v_\mathrm{bag}$.
Two cases are considered: the first is when the device is absorbing energy in power take-off (PTO) between the volumes V1 and V2 (see figure~\ref{sq1device}), and the second is when the device is forced to oscillate by harmonically varying V1 using a pump, in the absence of a PTO.
Note that V1 and V2 are used consistently throughout this paper to refer to the total volumes either side of the PTO or the pump, and that V1 can be greater or smaller than the physical volume of the bag, as given by~\eqref{bagvolume_static} below.

In the case where the device is absorbing energy in the PTO, the pressure amplitude $p_1$ and the volume amplitude $v_\mathrm{bag}$ are related according to
\begin{equation}
p_1 = -E v_\mathrm{bag} , \quad \text{where } \frac{1}{E} = \frac{V_2 C}{\gamma (P + P_\mathrm{atm})C + \mathrm{i}\omega M_2} + \frac{V_1}{\gamma (P + P_\mathrm{atm})} . \label{pcvcwave}
\end{equation}
The coefficient $C = \rho_\mathrm{air} / B_\mathrm{PTO}$, where $B_\mathrm{PTO}$ is the PTO damping, relates the mass flow through the PTO and the pressure difference across it:
\begin{equation}
\mathrm{i}\omega m_2 = -\mathrm{i}\omega m_1 = C (p_1 - p_2) . \label{C}
\end{equation}
Quantities corresponding to the volume V1 and the volume V2 have been denoted with subscripts 1 and 2, respectively.
Thus $V_1$ is the mean volume of air in V1, and $V_2$ is the mean volume of air in V2.
The corresponding mean masses of air in V1 and V2 are denoted by $M_1$ and $M_2$.
The air density at the mean pressure $P$ is  $\rho_\mathrm{air} = \rho_\mathrm{atm}  [(P + P_\mathrm{atm}) / P_\mathrm{atm}]^{1/\gamma}$, where $\rho_\mathrm{atm}$ is the air density at atmospheric pressure, $P_\mathrm{atm}$ is the atmospheric pressure, and $\gamma = 1.4$. 
Equation~\eqref{pcvcwave} has been derived using~\eqref{C} and considering the relevant linearised isentropic relations for V1 and V2:
\begin{equation}
\frac{v_\mathrm{bag}}{V_1} = \frac{m_1}{M_1} - \frac{p_1}{\gamma (P + P_\mathrm{atm})} 
\end{equation}  
for V1, and
\begin{equation}
p_2 = \gamma (P + P_\mathrm{atm}) \frac{m_2}{M_2}  \label{p2}
\end{equation}  
for V2.
Equation~\eqref{p2} can be combined with~\eqref{C} to give
\begin{equation}
-m_1 = m_2 = M_2 D p_1, \quad \text{where } D =  \frac{C}{\gamma (P + P_\mathrm{atm}) C + \mathrm{i}\omega M_2} . \label{D}
\end{equation}

In the case where the device is forced to oscillate by means of a pump in the absence of a PTO, the pressure amplitude $p_1$ and the volume amplitude of the bag $v_\mathrm{bag}$ are related more simply as
\begin{equation}
p_1 = -\gamma (P + P_\mathrm{atm}) \frac{v_\mathrm{pump} + v_\mathrm{bag}}{V_1} , \label{pcvcforced}
\end{equation}
where $v_\mathrm{pump}$ is the amplitude of the volume swept out by the pump. 

For both cases, the volume amplitude of the bag $v_\mathrm{bag}$ can be expressed in terms of the radial coordinates of the nodes on the tendon. 
The mean volume of the bag $V_\mathrm{bag}$ is first obtained as
\begin{equation}
V_\mathrm{bag} = \frac{\pi}{3} \sum_{n=1}^N (Z_n - Z_{n+1})(R_n^2 + R_n R_{n+1} + R_{n+1}^2)  \label{bagvolume_static}
\end{equation} 
by treating the bag as a stack of truncated cones.
The volume of the bag after displacement is then
\begin{multline}
V_\mathrm{bag} + v_\mathrm{bag} = \frac{\pi}{3} \sum_{n=1}^N (Z_n - Z_{n+1} + z_n - z_{n+1}) \\ \big[(R_n - r_n)^2 + (R_n + r_n) (R_{n+1} + r_{n+1}) + (R_{n+1} + r_{n+1})^2 \big] . \label{bagvolume_full}
\end{multline} 
Subtracting~\eqref{bagvolume_static} from~\eqref{bagvolume_full} and discarding higher-order terms yields the following expression for the bag volume amplitude:
\begin{multline}
v_\mathrm{bag} = \frac{\pi}{3} \sum_{n=1}^N \big\{(Z_n - Z_{n+1}) \left[ (2 R_n + R_{n+1} ) r_n + (R_n + 2 R_{n+1}) r_{n+1} \right] \\
+ (z_n - z_{n+1}) (R_n^2 + R_n R_{n+1} + R_{n+1}^2) \big\} .  \label{bagvolume_dynamic}
\end{multline} 

\subsubsection{Hydrodynamic forces} \label{hydro_forces}

The hydrodynamic pressures on the nodes $p_n^\mathrm{h}$ as well as the vertical hydrodynamic force on the ballast $f_\mathrm{B}^\mathrm{h}$, which appear in~\eqref{tendon_eq_dynamic} and~\eqref{ballast_eq_dynamic}, can be expressed as the sum of radiation and excitation components:
\begin{gather}
p_n^\mathrm{h} = \begin{cases}
0, &  n = 1,\dots,n_\mathrm{wl} - 1 \\
\sum_{k = n_\mathrm{wl}}^N (-r_k \sin A_{k-0.5} + z_k \cos A_{k-0.5}) p^\mathrm{R}_{n,k} + \xi_3 p^\mathrm{R}_{n,3} + p^\mathrm{exc}_n \label{p_hydro},  &  n = n_\mathrm{wl},\dots,N
\end{cases}  \\
f_\mathrm{B}^\mathrm{h} = \sum_{k = n_\mathrm{wl}}^N (-r_k \sin A_{k-0.5} + z_k \cos A_{k-0.5}) F^\mathrm{R}_{\mathrm{B},k} + \xi_3 F^\mathrm{R}_{\mathrm{B},3} + F^\mathrm{exc}_\mathrm{B} \label{F_hydro} .
\end{gather}
Here, $n = n_\mathrm{wl}$ is the first node below the waterline; 
$-r_k \sin A_{k-0.5} + z_k \cos A_{k-0.5}$ is the normal displacement of node $k$; 
$p^\mathrm{R}_{n,k}$ is the pressure on node $n$ due to a unit outward normal displacement of node $k$, i.e. of the lateral surface of a truncated cone with slanted height $h$ and centred on $k$; 
$p^\mathrm{R}_{n,3}$  is the pressure on node $n$ due to a unit heave displacement of the mean geometry; 
$p^\mathrm{exc}_n$ is the excitation pressure on node $n$; 
$F^\mathrm{R}_{\mathrm{B},k}$ is the vertical force on the ballast due to a unit normal displacement of node $k$; 
$F^\mathrm{R}_{\mathrm{B},3}$ is the vertical force on the ballast due to a unit heave displacement of the mean geometry; and
$F^\mathrm{exc}_\mathrm{B}$ is the vertical excitation force on the ballast.
The excitation pressure and force $p^\mathrm{exc}_n$ and $F^\mathrm{exc}_\mathrm{B}$ are zero for the pumped-oscillation case. 

%

%

\begin{figure}
\centering
\includegraphics[trim = 0mm 90mm 0mm 60mm, clip, scale=0.22]{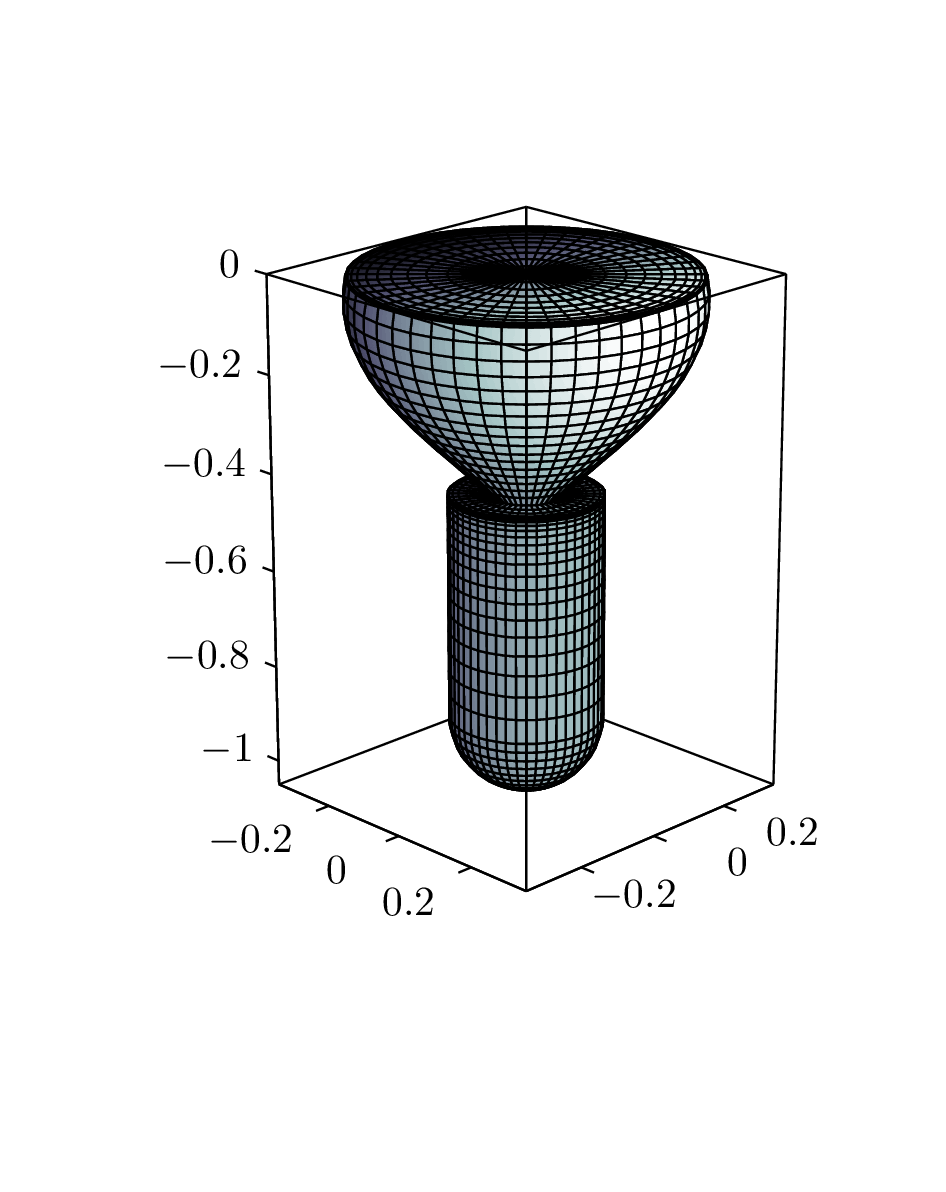} 
\caption{Typical panel model of the submerged geometry. Axis units are in metres.}
\label{panel_model}
\end{figure}

A panel method~\citep{WAMIT7.1} may be used to compute these hydrodynamic pressures and forces. 
The nodal pressures are the quantities of interest, so the bag is panelled such that nodes $n_\mathrm{wl}$ to $N$ as defined in figure~\ref{discretised_tendon} coincide with the panel centroids. 
A typical panel model is shown in figure~\ref{panel_model}
\footnote{In an alternative, self-contained, formulation of the problem, hydrodynamic pressures were computed by means of the semi-analytical theory of~\citet{Fenton1978} and~\citet{Isaacson1982} for wave forces on vertical axisymmetric bodies. 
For practical purposes the results were in all respects identical to those presented here.}.

It is helpful to note that 
\begin{gather}
-2\pi h  p^\mathrm{R}_{n,k} R_n = \iint_{S_\mathrm{b}} n_n p^\mathrm{R}_{n,k} \, \mathrm{d}S = F^\mathrm{R}_{n,k}  = \omega^2 A_{n,k} - \mathrm{i}\omega B_{n,k} , \\
F^\mathrm{R}_{\mathrm{B},k} = \iint_{S_\mathrm{b}} n_z p^\mathrm{R}_{\mathrm{B},k} \, \mathrm{d}S = \omega^2 A_{\mathrm{B},k} - \mathrm{i}\omega B_{\mathrm{B},k} ,
\end{gather}
where $S_\mathrm{b}$ is the wetted body surface, and thus by definition $A_{n,k}$ and $B_{n,k}$ are the added mass and radiation damping (in the normal direction) of node $n$ due to a unit normal displacement of node $k$, while $A_{\mathrm{B},k}$ and $B_{\mathrm{B},k}$ are the heave added mass and radiation damping of the ballast due to a unit normal displacement of node $k$. 
Likewise, 
$-2\pi h  p^\mathrm{R}_{n,3} R_n = \omega^2 A_{n,3} - \mathrm{i}\omega B_{n,3}$ and $F^\mathrm{R}_{\mathrm{B},3} = \omega^2 A_{\mathrm{B},3} - \mathrm{i}\omega B_{\mathrm{B},3}$.
Hence, 
the radiation forces may be expressed alternatively in terms of the appropriate added mass and radiation damping. 


\subsubsection{Final equations of motion and mean absorbed power}

We started by establishing the dynamic equations for the bag and the ballast,~\eqref{tendon_eq_dynamic} and~\eqref{ballast_eq_dynamic}. 
The complex amplitudes $p_1$, $p_n^\mathrm{h}$, $f_\mathrm{B}^\mathrm{h}$, $r_n$, $z_n$, $\xi_3$, $\tau$, and $a_n$ were the unknowns. 
By utilising the various relationships outlined in sections~\ref{radial-vertical_rel} to~\ref{hydro_forces}, we may express these unknowns in terms of only the radial displacements of the nodes $r_n$ and the heave displacement of the ballast $\xi_3$ as the independent unknowns.
By also making use of the boundary conditions $r_1 = r_{N+1} = a_1 = 0$, we finally arrive at a system of $N$ linear independent equations containing $r_2, r_3, \dots, r_N, \xi_3$ as the only unknowns. 
Such a system of equations can be written in matrix form as $\mathsfbi{Z} \boldsymbol{x} = \boldsymbol{F}$, where $\boldsymbol{x} = \begin{bmatrix} r_2 & r_3 & \cdots & r_N & \xi_3 \end{bmatrix}^\mathrm{T}$, and solved using standard methods. 


Once we obtain the displacements, we can calculate the mean absorbed power in the incident-wave case according to 
\begin{equation}
\mathcal{P} = \frac{C}{ 2 \rho_{\text{air}}} |p_1 - p_2|^2 ,
\end{equation}
where  the pressure amplitude $p_1$ can be obtained from~\eqref{pcvcwave} and~\eqref{bagvolume_dynamic}, and the pressure amplitude $p_2$ can be obtained from~\eqref{p2} and~\eqref{D}.

\section{Experiments}

Scaled model tests of the device were carried out in the 35 m $\times$ 15.5 m wave basin at Plymouth University, with a water depth of 3 m. 
The model tests were conducted at two different model sizes. 
The larger bag had a tendon length of 1.5 m and a displacement of 340 kg, while the smaller bag had a tendon length of 0.95 m and a displacement of 100 kg. 
Each bag had 16  tendons and the fabrics were both made of unreinforced polyurethane film. 
However, the construction of each of the bags was slightly different. 
The larger bag was made by joining together identical gores of fabric. 
The tendons, made of polyurethane-coated polyester strips, were welded along the common boundaries. 
The smaller bag was made from two circular fabric discs welded together along the perimeters. 
In this case the tendons, made of polyester-covered Vectran cores of 3-mm diameter, were not welded onto the fabric, but ran through guides attached to it.

The tendons terminated at the bottom of the bag at a radius of 15 cm for the larger bag and 7 cm for the smaller bag.
An aluminium disc inside the bag clamped the bottom of the bag to a steel cylinder which housed lead shot to provide the required displacement.
A hemispherical base was fitted to the bottom of the cylinder to minimise drag.
The larger cylinder had a radius of 203 mm and a height of 720 mm while the smaller cylinder had a radius of 152 mm and a height of 460 mm.
The mass of the cylinder including its contents was 454 kg for the larger model and 140 kg for the smaller model.



\begin{figure}
\centering
\includegraphics[trim = 3mm 0mm 0mm 0mm, clip, scale=0.9]{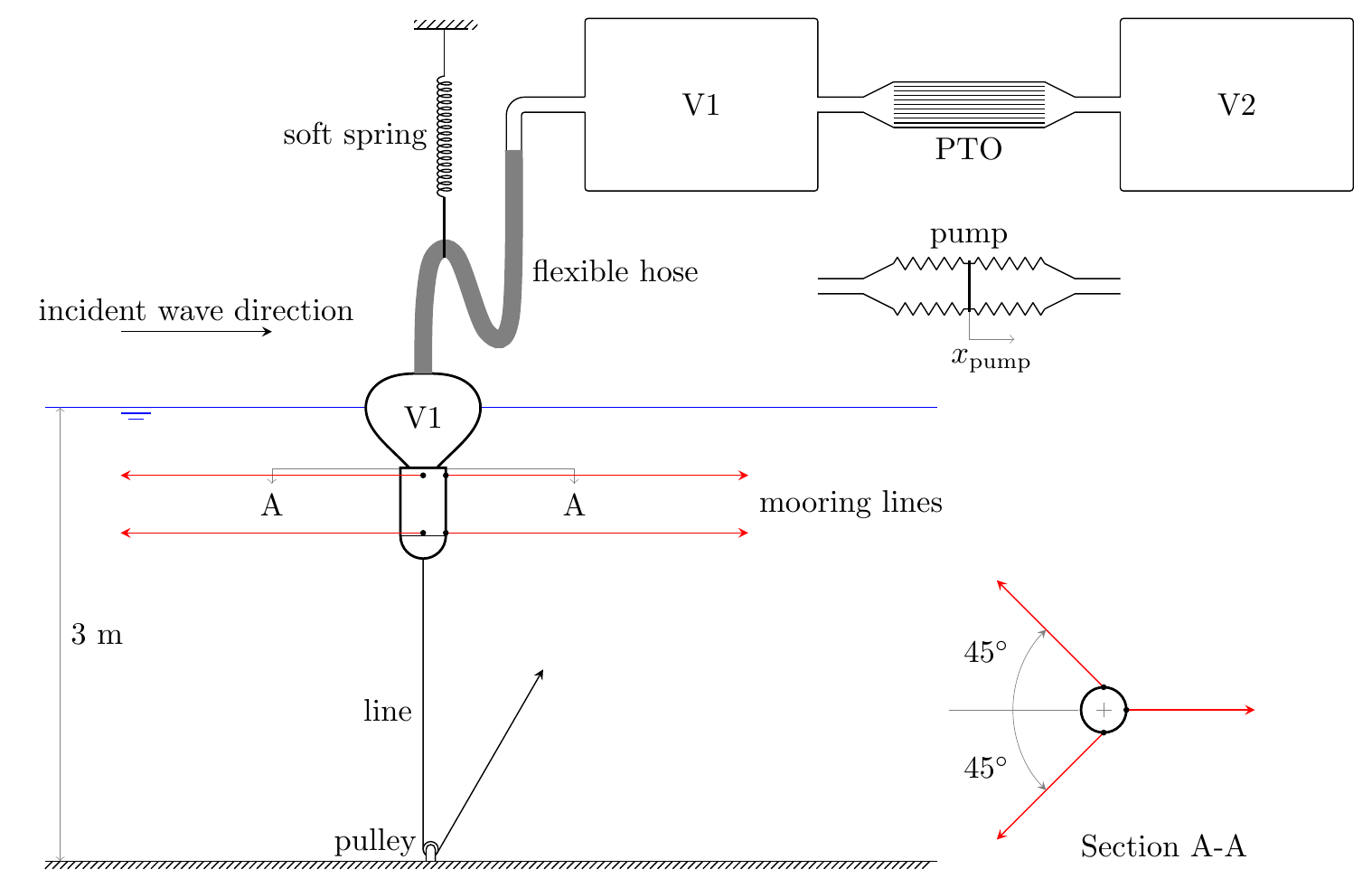}
\caption{Schematic of the experimental setup.
The line connected to the bottom of the ballast was used in the tests to measure the heave natural period of the device, but otherwise disconnected.}
\label{setup_side}
\end{figure}


Figure~\ref{sq1device} envisages a scheme in which V1 and V2 are contained within the device. 
If this arrangement were replicated at model scale $s$, the air stiffness and hence resonance heave frequency would be unrealistically high. 
In order to achieve an appropriate degree of compressibility of the air volumes, it was necessary to augment V1 and V2 with external sealed chambers as shown in figure~\ref{setup_side}.
The scale factor for V1 and V2 is 
\begin{equation}
\frac{V_\mathrm{model}}{V_\mathrm{prototype}} =  \frac{1}{s^3}\frac{P_\mathrm{atm} s + P_\mathrm{prototype}}{P_\mathrm{atm} + P_\mathrm{prototype}} . \label{air_volume_scaling}
\end{equation}
To allow free movement of the device, a flexible hose connected the bag to the additional volume.

For the larger model, one 1-m$^3$ tank was used to supplement the volume of the bag.
Therefore, $V_1$ was equal to the mean bag volume $V_\mathrm{bag}$ plus 1 m$^3$ plus the volume of the flexible hose and pipes on the V1 side. 
An identical 1-m$^3$ tank was used on the V2 side. 
For the smaller model, there was one 1.1-m$^3$ tank (in some cases closed off) on the V1 side and one or two 1.1-m$^3$ tanks were used on the V2 side.
These configurations are identified later as $0+1$, $1+1$, $1+2$, and so on. 
Either a PTO for the regular wave tests or a pump for the pumped-oscillation tests could be installed between V1 and V2.
The PTO was in the form of an assembly of parallel capillary pipes in which the air flow was laminar, providing a linear PTO of predictable damping~\citep{Chaplin2012}. 
This assembly consisted of 17 parallel tubes, each housing about 140 pipes of internal diameter 1.6 mm and length 800 mm.
The PTO damping could be varied according to the number of open tubes, where 9 was the minimum and 17 was the maximum. 
The covered range was between 73 and 39 kPa m$^{-3}$ s.

The oscillating pump consisted of two pairs of 300-mm diameter bellows on either side of diaphragms which were driven by electromagnetic digital linear actuators.
The pump moved such that any expansion of V1 is accompanied by a contraction of V2. 

For the smaller model, horizontal mooring lines with a configuration as shown in figure~\ref{setup_side} were used to limit pitch motions during the regular wave tests. 

Pressure transducers were used to record the pressures in V1 and V2. 
Two video cameras, one above and the other under water, recorded the motions of the device from the side. 
The vertical displacement of the top of the bag was recorded using a string potentiometer for the larger model, and infrared cameras for the smaller model. 
Other instrumentation included a manometer to monitor the pressure in the system at all times and an array of wave gauges to record the amplitude and phase of the incident waves in the regular wave tests.  


\section{Results and discussions} \label{results_and_discussions}

\subsection{Static behaviour} \label{static_results}

\begin{figure}
\centering
\includegraphics[trim = 1mm 1mm 0mm 0mm, clip, scale=0.88]{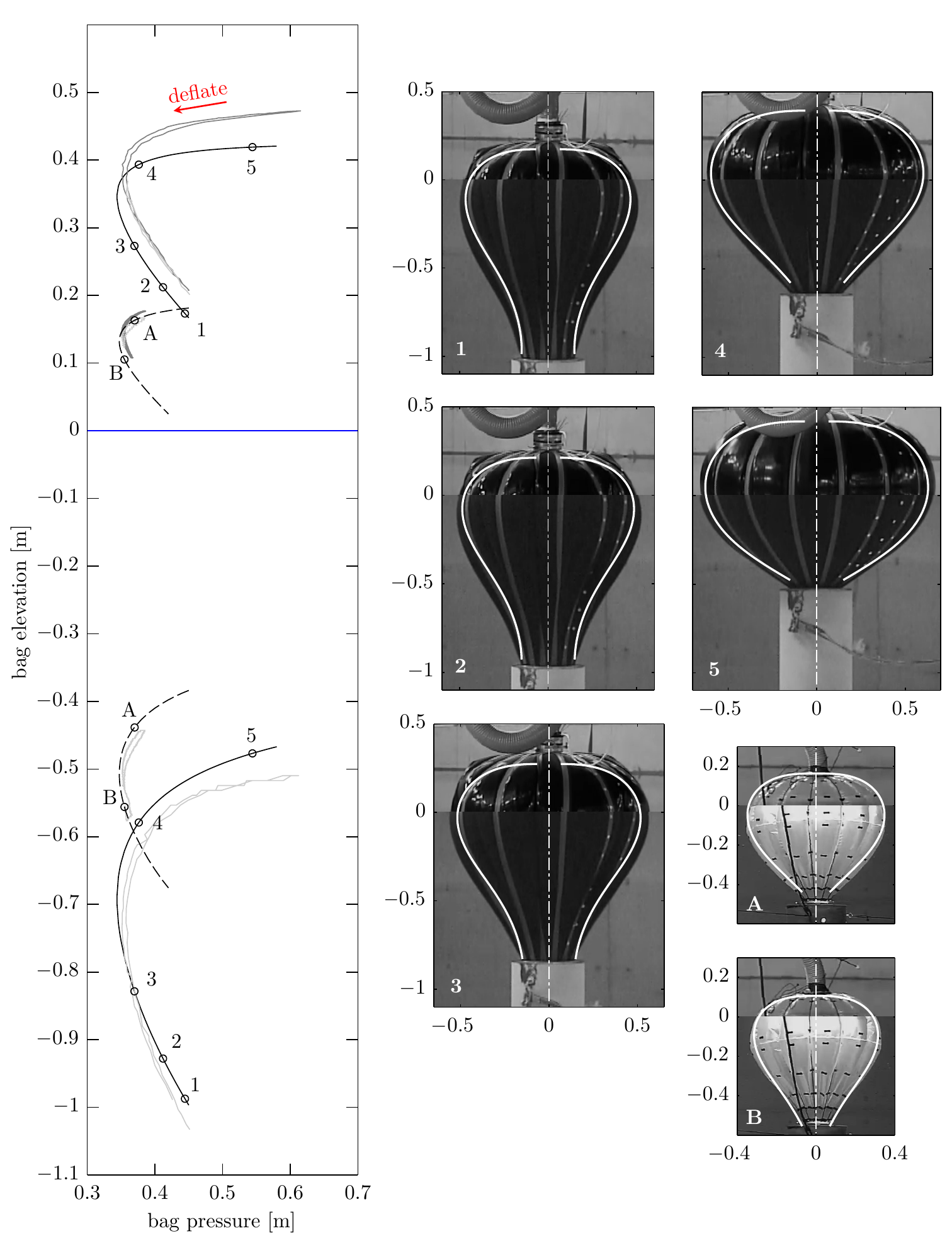} 
\caption{Static trajectories of the top and bottom elevations of the bag. Solid lines are the predicted trajectories of the larger bag. Dashed lines are the predicted trajectories of the smaller bag. Light grey lines are the measured trajectories obtained from video cameras. Dark grey lines are the measured trajectories obtained from string potentiometer for the larger bag, and from infrared cameras for the smaller bag. On the right are the predicted tendon profiles corresponding to the labelled points on the trajectories, superimposed on the recorded bag shapes at the same pressures. Axis units are in metres.}
\label{static_traj}
\end{figure}

The static behaviour of the bag floating in still water was investigated experimentally by measuring  
the elevations of the top and bottom of the bag and the internal pressure as the bag was inflated and deflated at a very slow rate. 
Plotting the top and bottom elevations versus the internal pressure of the bag results in C-shaped trajectories as shown in figure~\ref{static_traj}.
Some hysteresis can be observed in the measured trajectories, which indicates that the material did not immediately return to its original length as the bag was gradually deflated.
Following the trajectory, if we start somewhere near full inflation, i.e., on the upper side of the C curve, and deflate the bag slowly, the device would go down and the pressure would decrease until it reaches a minimum, and then the pressure would increase as the device continues to go down, before it sinks.
The trajectory therefore shows how the shape and pressure of the bag change as the mass of air in the system is increased or decreased.
The trajectory is a collection of all the possible combinations of elevation and pressure at which the device is in equilibrium.
Numerically, the trajectory is obtained by plotting the intersections found in figure~\ref{elevation-displacedvolume} as function of pressure. 
As predicted, depending on the amount of air in the bag, two different equilibrium bag shapes are possible which have the same pressure.


%


%

Figure~\ref{static_traj} shows, on the left, measured and predicted trajectories of the top and bottom of the bag. 
Seven points (five for the larger bag, cases 1 to 5, and two for the smaller bag, cases A and B) define the conditions for which images are shown on the right.
Some of these points also identify the initial conditions for subsequent measurements. 
The predicted bag shapes are about 10\% shallower than the actual shapes, as also seen from the images on the right. 
(Note that the outline of the bag in the recorded image is slightly larger than the actual tendon profile due to the fabric forming lobes between the tendons, and to a lesser extent, perspective effects.) 
The difference in height is due to the fact that in the physical bag some of the tension is shared by the fabric whereas in our calculations the tension in the fabric is neglected. 
With increasing number of tendons, this discrepancy is expected to become smaller.
Indeed, if tension in the fabric is accounted for,~\citet{Pagitz2010} have shown that for uniform internal-external pressure difference, the profile of a bag with a finite number of tendons is always higher than that of an isotensoid, i.e. the profile of the bag with infinitely many tendons. 

\subsection{Heave natural period}

\begin{figure}
\centering
\begin{overpic}[trim = 2mm 5mm 2mm 6mm, clip, scale=0.95]{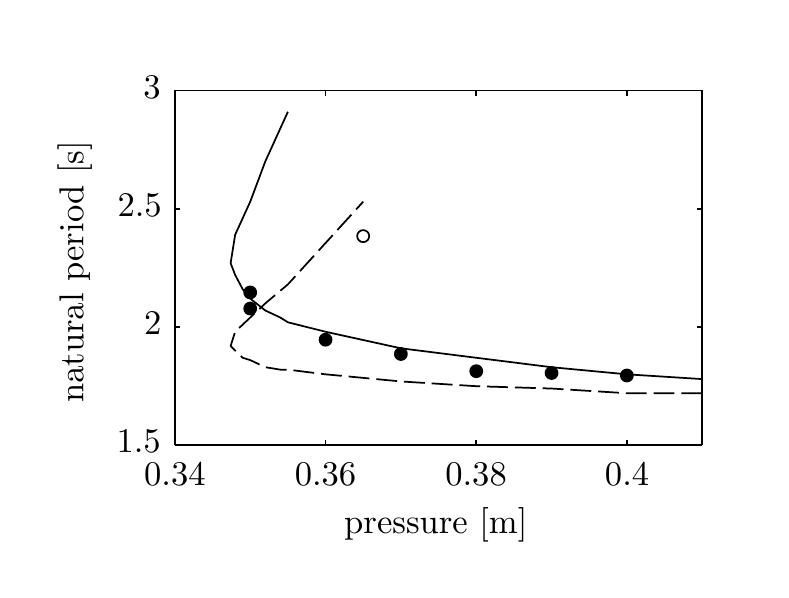}
\put(60,27){\begin{tikzpicture}[auto,scale=.5]
\draw [thick, -stealth, red] (0,0) -- node [above, sloped] {\footnotesize deflate} (-1.8,0.2);
\end{tikzpicture}}
\end{overpic}
\caption{Variation of heave natural period of the top of the bag with internal pressure, for the smaller model: numerical predictions (lines) vs. measurements (circles). Solid line and filled circles are with V1 + V2 tanks = 1 + 2. Dashed line and open circle are with V1 + V2 tanks = 0 + 2. The number of open PTO tubes is 17.}
\label{results_pressure-natperiod}
\end{figure}

By means of a line connected to the bottom of the ballast, as shown in figure~\ref{setup_side}, 
the device was displaced from its equilibrium position, and allowed to oscillate freely in still water. 
Natural periods in heave measured in this way for the smaller model (with all PTO tubes open) are shown in figure~\ref{results_pressure-natperiod} with corresponding predictions from the frequency-domain model.
The predicted natural period for each mean pressure was obtained from the peak period of the calculated heave response function of the top of the bag, when an oscillating vertical force was applied to the ballast, in otherwise still water.

The predictions and measurements are in reasonable agreement. 
Both show that  as  the bag is deflated the natural period lengthens, since its waterplane area gets smaller with decreasing freeboard.
Figure~\ref{results_pressure-natperiod} also shows the effect of the size of V1 on the device natural period. 
The effect is greater when the device is lower in the water. 


\begin{figure}
\centering
\includegraphics[trim = 0mm 3mm 0mm 5mm, clip, scale=0.95]{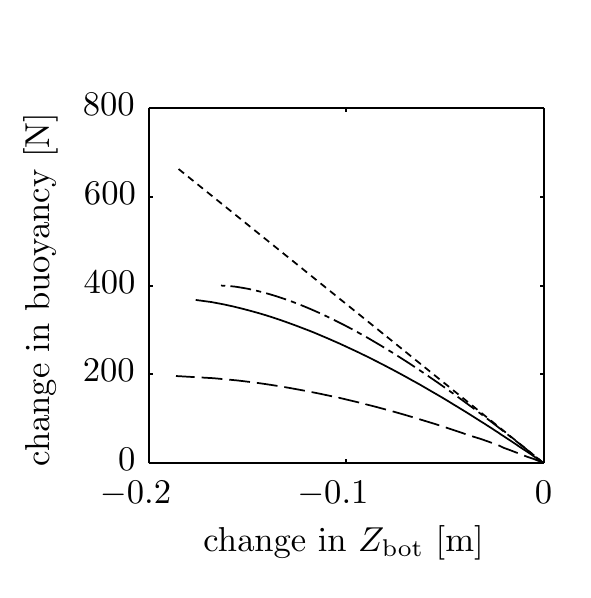}
\caption{Calculated change of buoyancy as the ballast is pulled down, for the smaller model with mean pressure of 37 cm (case A in figure~\ref{static_traj}) and $V_1$ equal to 0.18 m$^3$ (solid) and 2.41 m$^3$ (dashed); and for a rigid body of equal mean geometry (dash-dotted). Dotted line is the change in buoyancy if the water plane area is constant.}
\label{restoring}
\end{figure}

We can learn about the effect of V1 on the natural period also from static calculations by noting the change in buoyancy as the ballast is pulled down from its mean position.
Since the gradient of the curves in figure~\ref{restoring} is the hydrostatic restoring stiffness in heave, 
we see that  
with greater V1, the restoring stiffness is lower, and hence the natural period is longer. 
There is a critical air volume above which the floating bag device will be unstable. 

The restoring stiffness reduces with decreasing freeboard because the water plane area reduces at a greater rate when the waterline approaches the top of the bag. 
It is interesting to note that as the ballast is pulled down, a rigid body of equal mean geometry will get completely submerged sooner than the floating bag. 
This is due to the compressibility of the bag, which causes it to somewhat elongate vertically as the ballast moves down.


\subsection{Pumped oscillations}

In pumped-oscillation tests, harmonic pressure variations in the bag forced it to expand and contract, and the device to heave and radiate waves, in otherwise still water. 
In this case the secondary volume V2, as well as V1, varies with the pump amplitude. 
The pressure amplitude in V2, assuming its mean pressure equals that in the bag, is given as
\begin{equation}
p_2 = \gamma (P + P_\mathrm{atm}) \frac{v_\mathrm{pump}}{V_2} , \label{p2_forced}
\end{equation}
a function of the pump amplitude.
These tests were conducted with the larger model, for each of the five equilibrium conditions shown in figure~\ref{static_traj} (cases 1 to 5).
The heave response and the pressure variations were recorded. 

\begin{figure}
\centering
\includegraphics[trim = 30mm 117mm 30mm 18mm, clip, scale=.9]{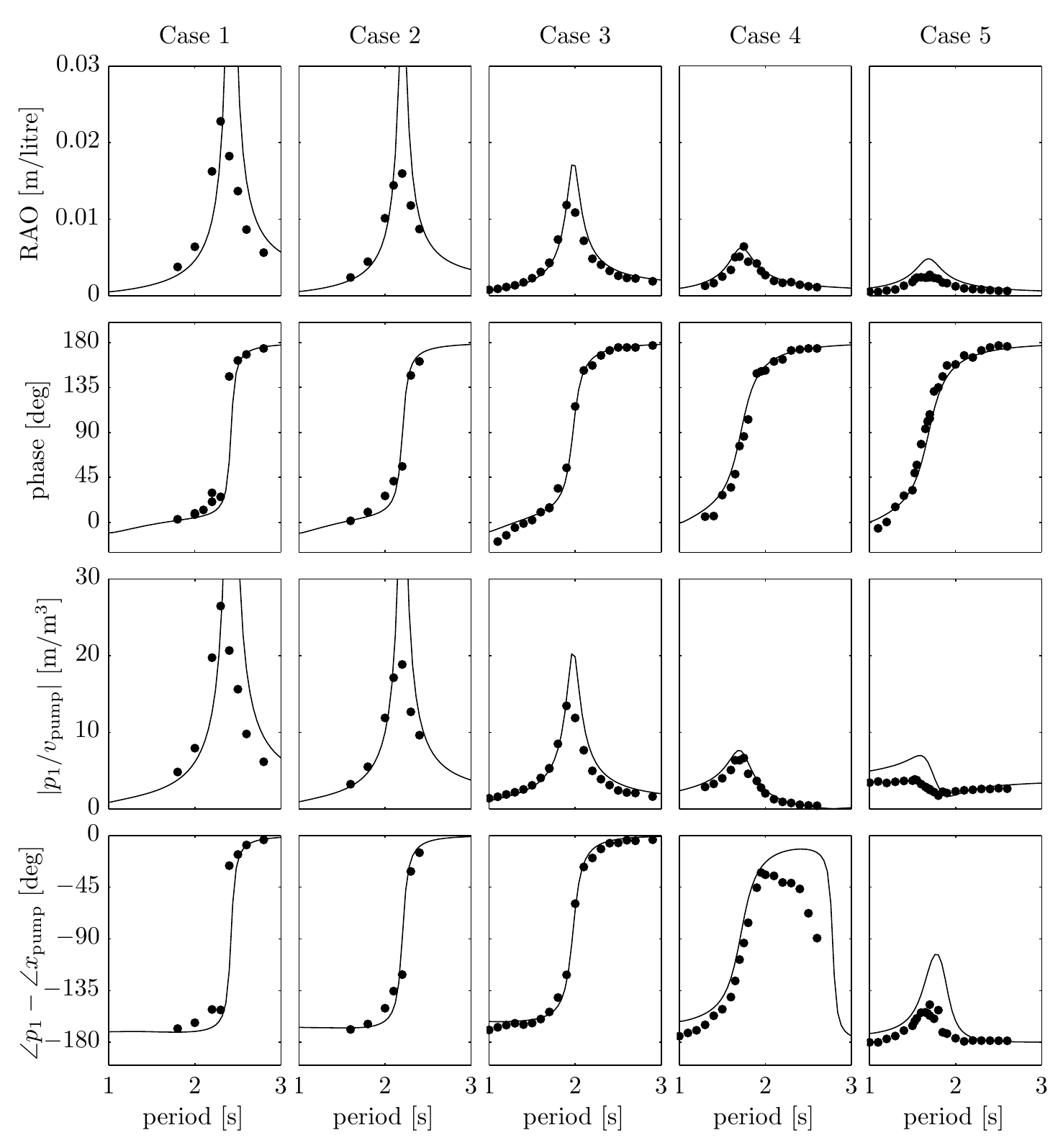}
\caption{Numerical predictions (lines) and experimental measurements (circles) of the response of the larger model when forced to oscillate by means of a pump, for cases 1 to 5 identified in figure~\ref{static_traj}. From top to bottom: heave displacement amplitude of the top of the bag per pump swept volume, phase of this displacement with respect to the pump displacement, pressure amplitude in V1 per pump swept volume, and phase of this pressure relative to the pump displacement. }
\label{results_pumped}
\end{figure}

Figure~\ref{results_pumped} shows, from top to bottom, the fundamental frequency components of the measured heave displacement of the top of the bag, phase of this displacement with respect to the pump displacement, pressure amplitude in the bag, and the phase of this pressure relative to the pump displacement, along with the numerical predictions.  

In general, the predictions are in good agreement with the measurements, with most features predicted reasonably well, such as the lengthening of resonance period from case 5 to 1 and the accompanied increase in peak amplitudes of both the heave displacement and the bag pressure. 
The phase variations are also predicted reasonably well. 
The predicted resonance periods are slightly longer than the measured ones for all cases, except for case 4. 
This is generally due to the physical bag having a slightly larger water plane area due to the presence of the lobes.
The predicted peak amplitudes are also slightly higher, except again for case 4, where the predicted amplitudes match well with the measured ones. 
The relatively better agreement between the predicted and measured peak amplitudes in case 4 (and to some extent case 3) compared to that in the other cases is probably due to the fact that the shape of the gores for this larger bag was designed based on a static tendon profile calculated at mean pressure close to those in cases 4 and 3. 


It is worth noting that in
cases 4 and 5, 
the pressure amplitude goes to a minimum at some period longer than the resonance period, and there is a phase shift associated with this minimum. 
At this period, the volume swept by the pump is cancelled by the volume change of the bag such that the total volume remains approximately constant.   
For cases 1 to 3 this probably happens at some period beyond the range considered in the tests.

\subsection{Incident wave excitations}

Regular wave tests were conducted over a wide range of conditions with the smaller model. 
The incident wave amplitude was always 2 cm.

\begin{figure}
\centering
\includegraphics[trim = 38mm 145mm 37.5mm 16mm, clip, scale=1]{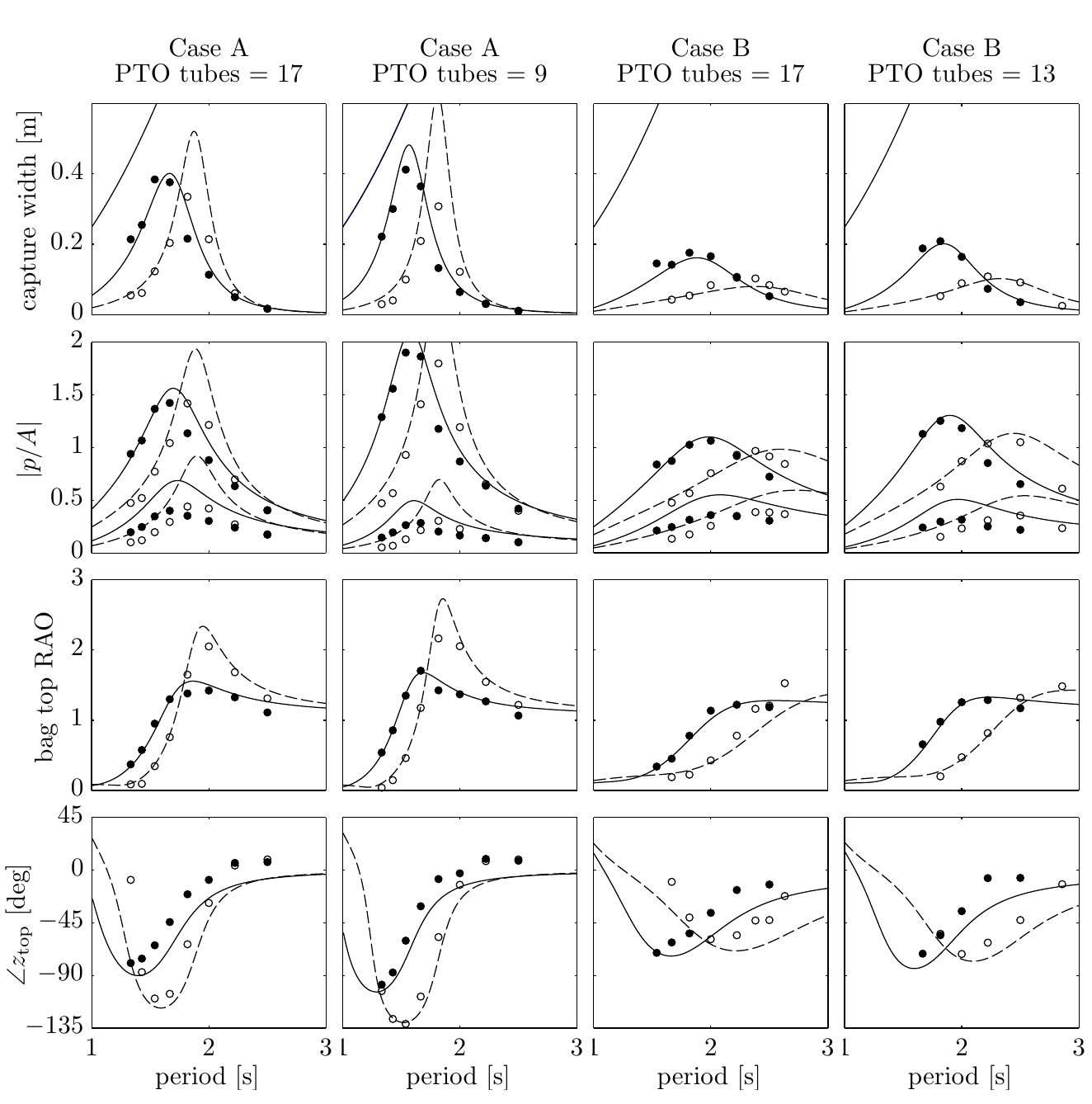} 
\caption{Numerical predictions (lines) and experimental measurements (circles) of the response of the smaller model in regular waves. 
From top to bottom: capture width, pressure amplitudes in V1 and V2 per incident wave amplitude ($p_1$ being the higher ones), heave displacement amplitude of the top of the bag per incident wave amplitude, and phase of this displacement with respect to the incident wave elevation at origin. 
Solid lines and filled circles correspond to V1 + V2 tanks = 0 + 2, while dashed lines and open circles correspond to V1 + V2 tanks = 1 + 2. 
The theoretical capture width limit $\lambda/2\pi$ is shown as the ascending solid lines in the capture width plots.
The case as identified in figure~\ref{static_traj} and the number of open PTO tubes are indicated at the top of each column.}
\label{results_wave}
\end{figure}

Figure~\ref{results_wave} shows fundamental frequency components of the measured response of the device along with the numerical predictions.
Plotted alongside the measured and predicted capture widths is the well-known theoretical limit $\lambda / 2 \pi$, where $\lambda$ is the incident wave length, applicable for point absorbers or source-mode radiators~\citep{Budal1975,Evans1976,Newman1976},
to which category the present device belongs.

On the whole, the agreement is good. 
Some discrepancies can be observed between the measured and predicted pressure amplitude $p_2$, where the predictions are seen to overestimate the measurements. 
This is probably due to uncertainties associated with the actual volumes of the tanks and the fact that we have neglected any other damping apart from the
PTO damping 
and the wave radiation damping. 
Some discrepancies are also evident on the phase of the bag top displacement. 
These may be attributed to uncertainties in the distance between the wave gauges and the device axis, since the mean position of the device varied slightly from one test to another. 


From the figure we observe that increasing $V_1$ has the effect of lengthening the heave resonance period of the device.
The capture width curve is shifted to longer periods, which is an advantage compared to a rigid device of equal size. 
In section~\ref{parametric_study} we will show that quite a significant increase in resonance period can be achieved without making $V_1$ unrealistically large.
%
Comparing the responses between cases A and B, we see that the heave resonance period is also lengthened by bag deflation, since the mean geometry changes with deflation. 
%
%
%
Figure~\ref{results_wave} also shows that having more PTO tubes open, i.e. decreasing the PTO damping, reduces the pressure amplitude in V1 and increases the pressure amplitude in V2.
This is reasonable, as with more PTO tubes open, the two volumes become more like a single volume and the difference between the two pressures gets smaller. 
Indeed, based on the linearised isentropic relation, the ratio of $p_1$ to $p_2$ can be shown to be
\begin{equation}
\frac{p_1}{p_2} = 1 + \frac{\mathrm{i} \omega V_2 \rho_\mathrm{air}}{C \gamma (P + P_\mathrm{atm})} ,
\label{pressureeq}
\end{equation}
where $\omega$ is the wave frequency.
For the same mean pressure $P$, we see that as $V_2 / C$ gets smaller, the pressures in the two volumes, $p_1$ and $p_2$, become more similar both in amplitude and phase.
On the other hand, as $V_2 / C$ gets larger, the pressure amplitude $p_2$ tends to zero while the phase difference between the two pressures tends to 90$^\circ$.

An interesting detail is observed in the heave displacement of the top of the bag at short periods, especially for the case of 37 cm mean pressure, 9 PTO tubes, and V1 + V2 tanks = $1+2$, where it is most clearly seen that the response has a minimum at about 1.2 s (this is also seen to some extent in the other three cases, but the relatively more significant damping makes the response rather flat here).
There is a phase shift associated with this minimum, meaning that as the ballast moves up, the top of the bag moves down relative to the ballast such that the total displacement $z_\mathrm{top} = z_1 + \xi_3$ is minimum. 
Such behaviour is not observed when the bag is forced to oscillate in still water (see figure~\ref{results_pumped}), suggesting that this is specific to the incident wave case.

\begin{figure}
\centering
\includegraphics[trim = 32mm 232mm 32mm 19mm, clip, scale=0.9]{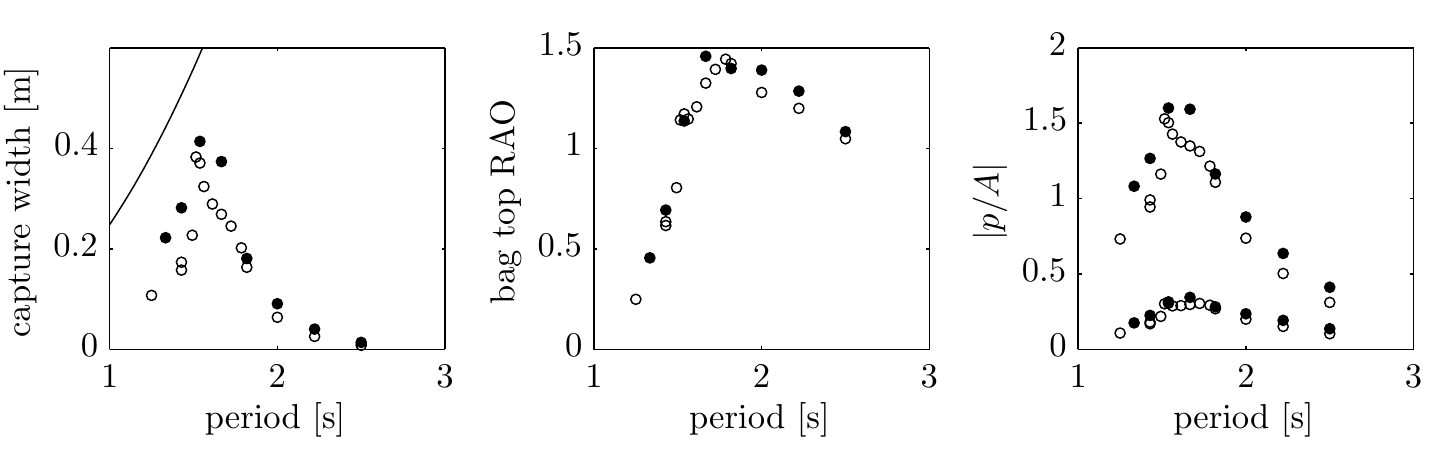} 
\caption{Measured capture widths, heave displacement amplitude of the top of the bag per incident wave amplitude, and pressure amplitudes in V1 and V2 per incident wave amplitude ($p_1$ being the higher ones), with the device pitch restrained (filled circles) and free (open circles), for case A in figure~\ref{static_traj}, with V1 + V2 tanks = $0+2$ and 13 PTO tubes open.
The theoretical capture width limit $\lambda/2\pi$ is shown as the ascending line in the capture width plot.}
\label{results_slack}
\end{figure}

Allowing the device to pitch more freely by releasing tension in the mooring lines is found to decrease the absorbed power, pressure amplitudes, and heave response of the bag (see figure~\ref{results_slack}), but the effects appear to be small.  
Since heave and pitch should be uncoupled for a body with fore-aft symmetry according to linear theory, the decrease is likely due to some nonlinear coupling between heave and pitch. 
When the device pitches more, it is found to heave less, and this leads to a decrease in the pressure amplitudes in the bag. 
With the same PTO damping, this in turn results in a decrease in the absorbed power. 
Similar phenomenon has been observed by~\citet{Gomes2015b,Gomes2015} for an axisymmetric floating oscillating-water-column device.

\subsection{Parametric analysis} \label{parametric_study}

With the numerical model validated (at least for conditions where linear assumptions are expected to be valid), we will now examine the response of the device under a wider range of parameters.

\begin{figure}
\centering
\includegraphics[trim = 33mm 75mm 33mm 17mm, clip, scale=0.9]{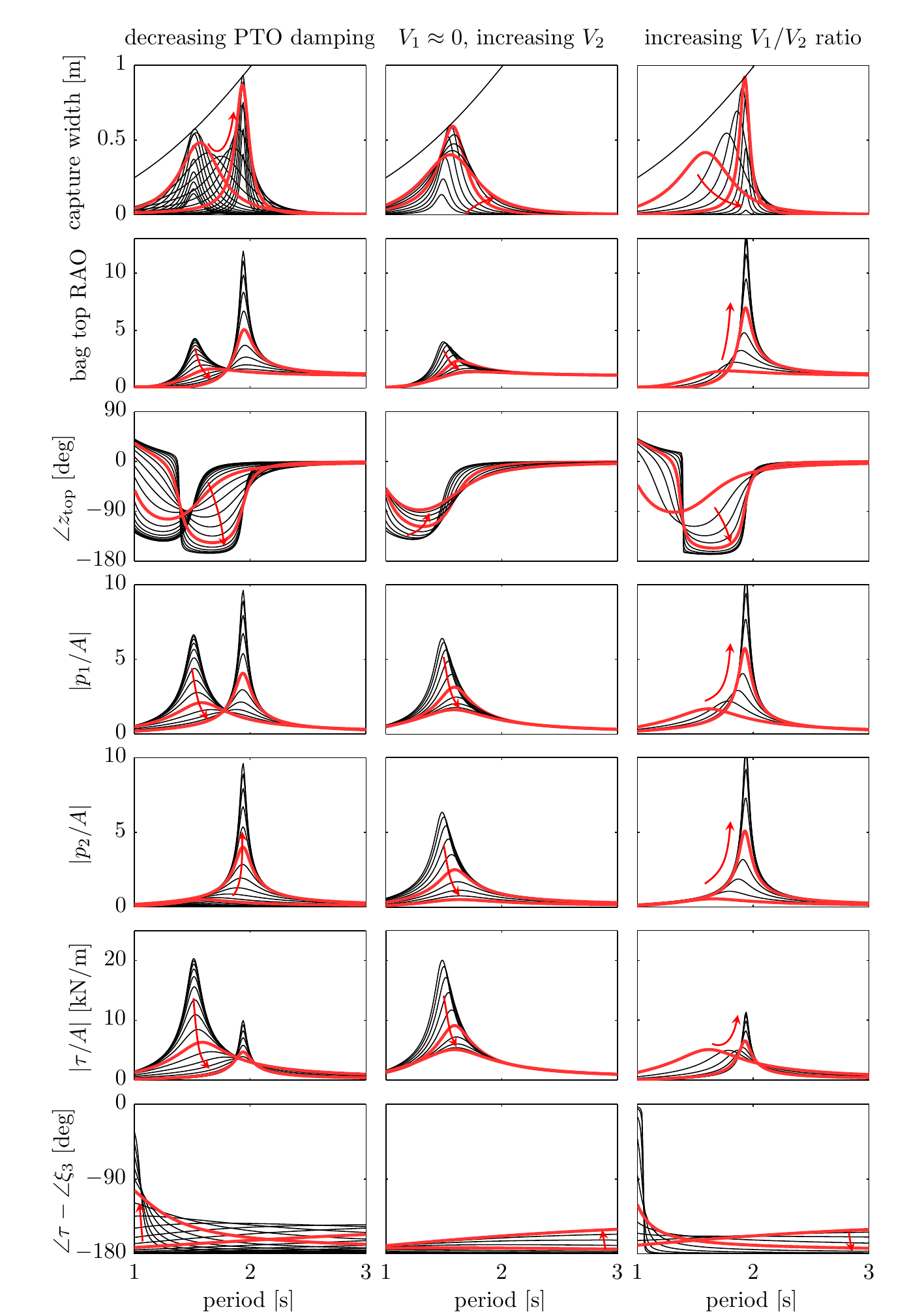}
\caption{Left column: Effects of varying PTO damping, with $V_1 = 0.18$ m$^3$ and $V_2 = 2.23$ m$^3$. 
Arrow indicates decreasing PTO damping. 
Thicker lines indicate responses with 9 and 180 PTO tubes. 
Middle column: Effects of varying $V_2$, with $V_1 \approx 0$ and 13 PTO tubes. 
Arrow indicates increasing $V_2$.
Thicker lines indicate responses with $V_2 = 0.59$ m$^3$ and $V_2 = 2.41$ m$^3$.
Right column: Effects of varying $V_1/V_2$, with $V_1+V_2 = 2.41$ m$^3$ and 13 PTO tubes.
Arrow indicates increasing $V_1/V_2$. 
Thicker lines indicate responses with $V_1/V_2 = 0.04$ and $V_1/V_2 = 3.96$.
All for case A in figure~\ref{static_traj}.
The theoretical capture width limit $\lambda/2\pi$ is shown as the ascending solid lines in the capture width plots.}
\label{var_PTO}
\end{figure}

The left column of figure~\ref{var_PTO} shows the effects of varying the PTO damping on the response of the smaller device with all other parameters unchanged. 
For very large or very small PTO damping, only little power is absorbed. 
Between these two extremes, the peak capture width gradually increases and then decreases, before increasing and decreasing again, touching the theoretical limit twice as the peak shifts from shorter to longer periods, resulting in a double-peaked envelope. 
Note that each capture width curve is single-peaked, only the envelope is double-peaked. 
The heave displacement, pressure, and tension amplitudes are characterised by the same shift in peak period, but the peak amplitudes are the highest in the limit of very large or very small PTO damping.
At the trough of the envelope, the response, including the capture width, is most broad-banded. 

Such a double-peaked envelope appears to be a characteristic of devices in which air is exchanged between two closed volumes through a PTO~\citep{Kurniawan2014a}.
We can show that in the limit of very large or very small PTO damping ($C \to 0$ or $C \to \infty$, respectively), the pressure amplitudes $p_1$ and $p_2$ become independent of the PTO damping. 
From~\eqref{pcvcwave},~\eqref{p2}, and~\eqref{D}, we have
\begin{equation}
\lim_{C \to 0} p_1 = - \frac{\gamma (P + P_\mathrm{atm})}{V_1} v_\mathrm{bag} \label{p1_C=0} \quad \text{and} \quad
\lim_{C \to 0} p_2 = 0 
\end{equation}
 for very large PTO damping ($C \to 0$), whereas for very small PTO damping ($C \to \infty$),
\begin{equation}
\lim_{C \to \infty} p_1 = - \frac{\gamma (P + P_\mathrm{atm})}{V_1 + V_2} v_\mathrm{bag} \quad \text{and} \quad
\lim_{C \to \infty} p_2 = p_1 .
\end{equation}
In other words, when $C \to 0$, the secondary volume V2 becomes practically sealed from V1, and the pressure amplitude in V1 becomes what we would have if there were only one volume V1 and no PTOs (note that~\eqref{p1_C=0} is the same as~\eqref{pcvcforced} and~\eqref{p2_forced} with $v_\mathrm{pump} = 0$).
Likewise, when $C \to \infty$, the two volumes V1 and V2 effectively combine into one single volume and the pressures in the two volumes become equal. 
In both of these limiting cases, the absorbed power tends to zero.
The first peak in the plots on the left column of figure~\ref{var_PTO} is associated with the heave resonance of a ballasted floating bag (without PTO) with air volume equal to $V_1$, while the second peak with air volume $V_1 + V_2$.
The separation between the two peaks in the envelope therefore is related to the difference between $V_1$ and $V_1+V_2$. 
Increasing $V_2$ while keeping $V_1$ constant will increase the separation, but the trough between the two peaks will also be deeper. 
This implies that to capture appreciable power at periods between the two peaks, $V_1$ must also be adjusted in addition to the PTO damping.

The middle and right columns of figure~\ref{var_PTO} show the effects of varying $V_1$ and $V_2$ with constant PTO damping. 
It is assumed that the total available volume $V_1+V_2$ is 2.41 m$^3$, as in the experiments with V1 + V2 tanks = $1 + 1$. 
For the middle column, $V_1$ is small and kept constant while $V_2$ is varied. 
For the right column, both $V_1$ and $V_2$ are varied while the sum of $V_1+V_2 = 2.41$ m$^3$ is maintained. 
We can show that when $V_1 \to 0$ (regardless of $V_2$), 
\begin{equation}
\lim_{V_1 \to 0} m_1 = \rho v_\mathrm{bag},
\end{equation}
that is, all the air that is compressed by the bag will flow through the PTO, since the air compressibility becomes negligible when $V_1$ is small. 
When $V_2 \to 0$ (regardless of $V_1$), 
\begin{equation}
\lim_{V_2 \to 0} p_1 = - \frac{\gamma (P + P_\mathrm{atm})}{V_1} v_\mathrm{bag}  \quad \text{and }
\lim_{V_2 \to 0} p_2 = p_1 ,
\end{equation}
and the absorbed power tends to zero. 
On the other hand, when $V_2 \to \infty$ (regardless of $V_1$), 
\begin{equation}
\lim_{V_2 \to 0} p_2 = 0 .
\end{equation}
So decreasing $V_2$ and decreasing the PTO damping have similar effects on $p_2$ but opposite effects on $p_1$. 


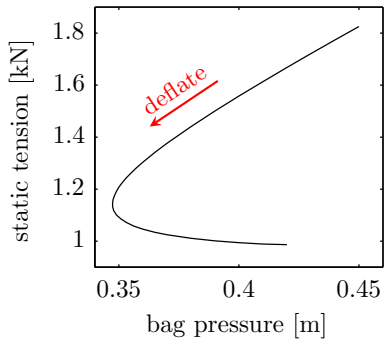
\begin{figure}
\centering
\begin{overpic}[trim = 0mm 3mm 0mm 5mm, clip, scale=0.95]{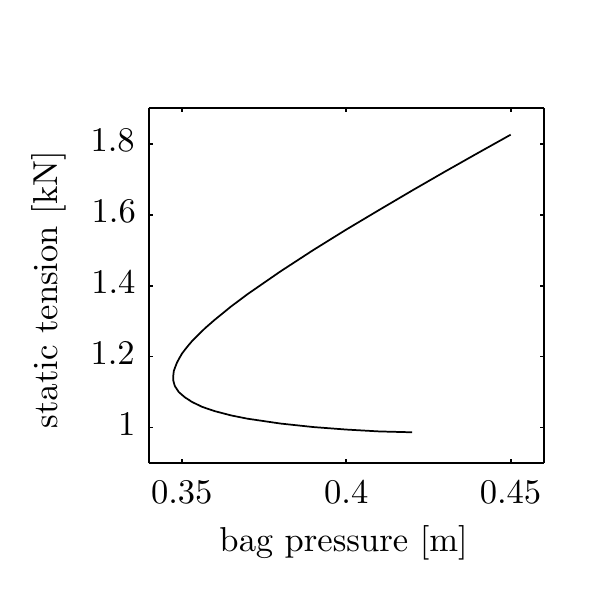}
\put(33,50){\begin{tikzpicture}[auto,scale=.5]
\draw [thick, -stealth, red] (0,0) -- node [above, sloped] {\footnotesize deflate} (-1.8,-1.2);
\end{tikzpicture}}
\end{overpic}
\caption{Variation of static tension in all tendons with mean internal pressure, for the smaller model.}
\label{tension_traj}
\end{figure}

\begin{figure}
\centering
\includegraphics[trim = 31.5mm 190mm 32mm 18mm, clip, scale=.95]{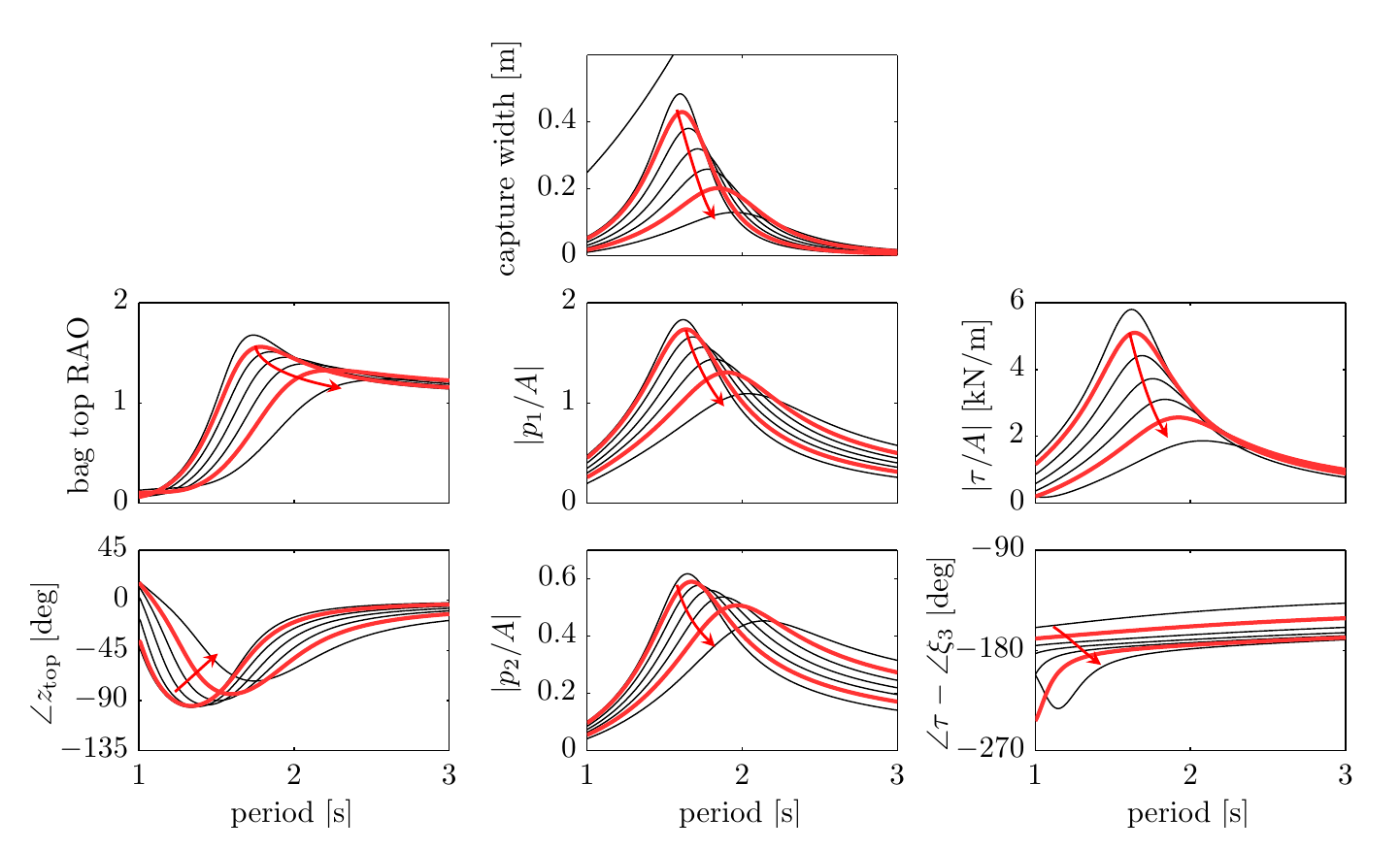} 
\caption{Effects of varying mean pressure on the response of the smaller model with $V_1 = 0.18$ m$^3$, $V_2 = 2.23$ m$^3$, and 13 PTO tubes. 
Arrow indicates how the response varies with decreasing amount of air (descend along the static trajectory). 
Thicker lines indicate responses for cases A and B identified in figure~\ref{static_traj}. 
The theoretical capture width limit $\lambda/2\pi$ is shown as the ascending line in the capture width plot.}
\label{var_P}
\end{figure}

Figure~\ref{tension_traj} shows the variation of the predicted static tension in all tendons with the internal pressure of the bag. 
On the upper side of the trajectory, the static tension varies almost linearly with bag pressure. 
From figure~\ref{var_PTO} we see that in the dynamic case the amplitude of tension in the tendons also appears to correlate with the amplitude of pressure in the bag. 
Furthermore, the tension is found to be maximum when the ballast is near its lowest position, except at very low periods where a phase reversal happens and the tension becomes more in phase with the ballast. 
Figure~\ref{var_PTO} shows that the tension amplitude can be controlled by adjusting the PTO damping or the volume ratio. 
Another way to reduce the tension amplitude is by deflating the bag, as shown in figure~\ref{var_P}.
However, there is usually a trade-off between minimising loads and maximising the absorbed power, and this is the case here as well.

\begin{figure}
\centering
\includegraphics[trim = 35mm 200mm 30mm 19mm, clip, scale=0.95]{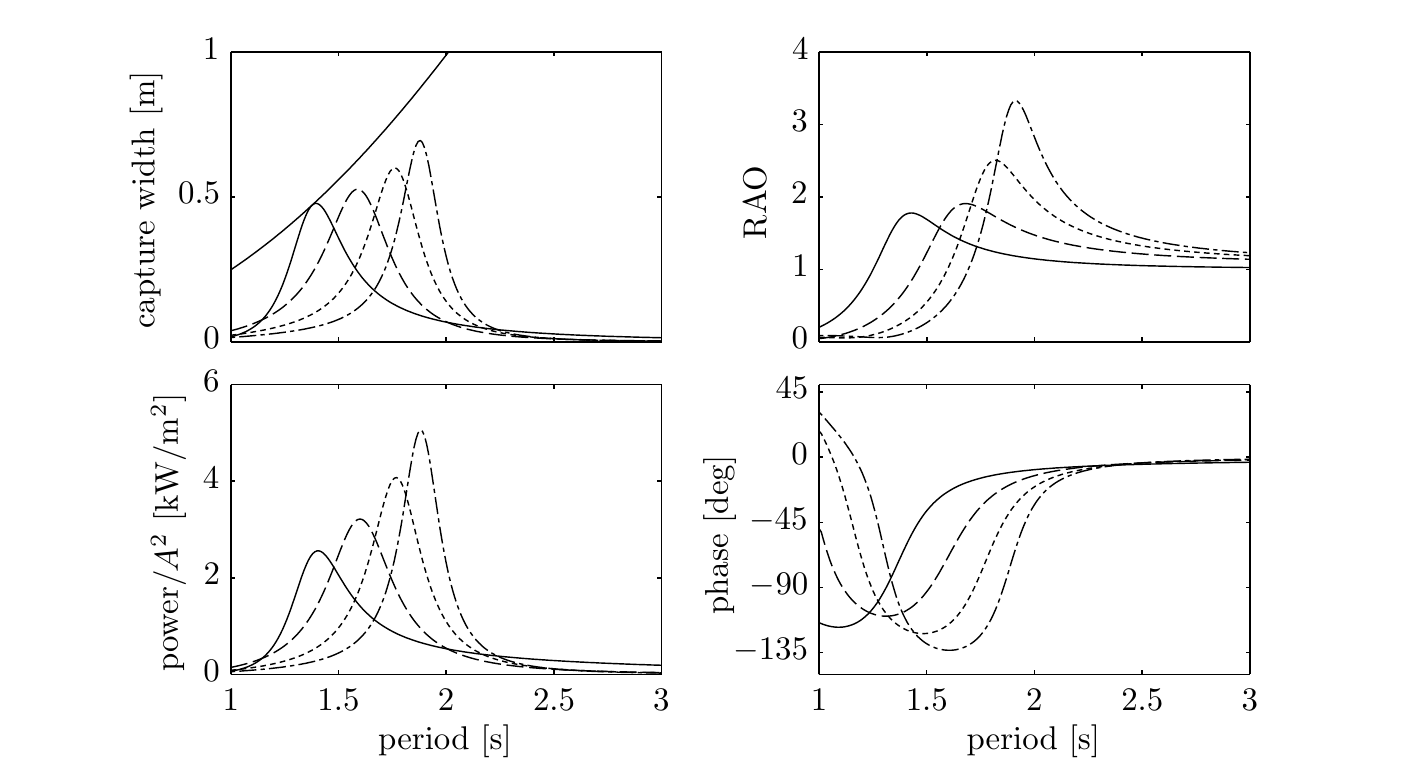} 
\caption{Response comparison of the smaller model of the floating bag device in case A, with $V_2 = 1.13$ m$^3$ and various $V_1$ and PTO settings: $V_1 = 0.18$ m$^3$ and 9 open PTO tubes (dashed); $V_1 = 0.73$ m$^3$ and 13 open PTO tubes  (dotted); $V_1 = 1.28$ m$^3$ and 17 open PTO tubes (dash-dotted); and a rigid body of equal mean geometry absorbing power through heave, with PTO damping equal to radiation damping at resonance (solid).
The theoretical capture width limit $\lambda/2\pi$ is shown as the ascending line in the capture width plot. 
For the floating bag device, the RAOs and the phases are those of the top of the bag. The RAOs of the ballast are somewhat higher.}
\label{rigid_comp}
\end{figure}

Figure~\ref{rigid_comp} compares the response of the smaller model of the floating bag device in case A to that of a rigid body of equal mean geometry, which is absorbing power through heave. 
The parameters of the floating bag device are 
selected for 
illustration only. 
Clearly, as the resonance period of the floating bag device is higher than that of the rigid device, the floating bag device can capture more energy from longer waves. 

\begin{figure}
\centering
\includegraphics[trim = 31mm 232mm 30mm 18mm, clip, scale=0.9]{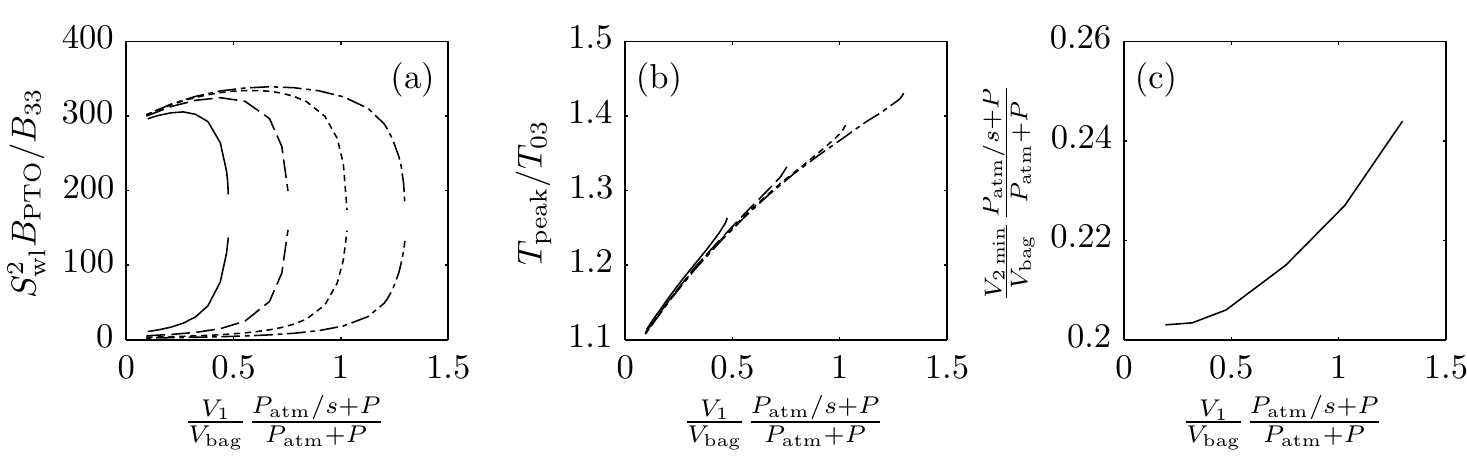} 
\caption{(a) Optimum normalised PTO damping as a function of normalised V1 volume, for a floating bag device 
where $s = 25$ corresponds to case A in figure~\ref{static_traj}.
$S_\mathrm{wl}$ is the waterplane area and $B_{33}$ is the resonant heave radiation damping of a rigid body with equal mean geometry.
(b) Corresponding peak period of the absorbed power relative to the heave natural period of a rigid body with equal mean geometry.
For (a) and (b), different line types indicate different values of $\frac{V_1 + V_2}{ V_\mathrm{bag}} \frac{P_\mathrm{atm}/s + P}{P_\mathrm{atm} + P}$: 0.682 (solid), 0.969 (dashed), 1.256 (dotted),  1.543 (dash-dotted).
(c) Minimum $V_2$ (normalised) for the absorbed power to reach the theoretical maximum.}
\label{design}
\end{figure}

If we adopt a scale of $s = 25$ for the smaller model, the total displacement of the device will be about 2200 tonnes at full scale.
The full-scale waterplane diameter of the bag will be about 17 m and the bag volume 2200 m$^3$ at a full-scale mean pressure of 0.9 bar. 
The required total air volumes corresponding to the three cases in figure~\ref{rigid_comp} will be 1498 m$^3$, 2127 m$^3$, and 2756 m$^3$ at full scale, and the corresponding peak periods of the absorbed power will be 8~s, 8.85~s, and 9.4~s. 
The latter is over 30\% longer than the peak period of the full-scale rigid device. 

Figure~\ref{design}a shows what values the PTO damping must take to maximise the absorbed power at resonance. 
These optimum dampings are plotted in figure~\ref{design}a as functions of $V_1$, for various total $V_1 + V_2$ volumes.
There are two optimum damping values for each $V_1$, the larger damping giving the first peak of the envelope described earlier, and the smaller one giving the second peak.
The variation of the first peak period of the absorbed power is plotted in figure~\ref{design}b.
 
It is worth noting that in order for the absorbed power to reach the theoretical maximum, $V_2$ must be greater than a certain minimum which depends on $V_1$. 
As seen from figure~\ref{design}c, the required minimum $V_2$ at full scale is less than half the bag volume.

\section{Conclusion}

Numerical and experimental investigations have been carried out on a novel wave energy device incorporating a floating air bag. 
The bag expands and contracts in waves, thus creating a reciprocating air flow through a turbine between the bag and a secondary volume.
The dynamic properties of the device can be altered either by varying the amount of air in the bag, the turbine damping, or the ratio between the primary volume and the secondary volume. 
These offer relatively simple means to adapt the device to various wave conditions.

A method of calculating the static shape of the bag for a given combination of internal pressure and submerged weight of the ballast has been presented, where 
it is shown that depending on the amount of air in the system, two different bag shapes can have the same internal pressure.   
An approach for modelling the device dynamics under forced-oscillations or in incident waves has also been described, and the results agree closely with measurements from model tests. 
The challenge of ensuring correct compressibility of the bag at small scale has been overcome in the tests by connecting the bag via a flexible hose to some appropriately scaled additional air volumes. 

As expected, the resonance period of the device is longer than that of a rigid device with the same geometry, even up to 30\% longer with air volumes that can all be contained within the device itself at full scale. 
There is an accompanied loss in bandwidth compared to the bandwidth of the rigid device, but the loss is not severe.
In a typical sea state, the floating bag device can capture about twice as much energy as a rigid device of the same volume. 

All the analysis presented in this paper assumes small displacements for the adopted linear theory to be valid, and all the incident wave tests have been conducted with small-amplitude waves.  
The deviation from linearity is evident for example from figure~\ref{restoring}. 
When the excursion of the ballast is about 6 cm for the smaller model, the expected deviation from linearity is about 5\%.
The deviation however increases to about 20\% when the excursion is about 10 cm.   
To account for such nonlinearity, a nonlinear time-domain model would be necessary.
A time-domain model will also be useful to simulate the application of phase control techniques, such as latching, which may be appropriate for this type of device.   \\

This work is supported by the EPSRC SuperGen Marine Energy Research Consortium [EP/K012177/1]. 
The authors are grateful to Prof.~Francis Farley for stimulating discussions and to Malcolm Cox of Griffon Hoverwork Ltd for supplying the model-scale bags and for practical insights.

\bibliographystyle{jfm}
\bibliography{wavenergy}{}

\begin{thebibliography}{32}
\expandafter\ifx\csname natexlab\endcsname\relax\def\natexlab#1{#1}\fi
\def\au#1{#1} \def\ed#1{#1} \def\yr#1{#1}\def\at#1{#1}\def\jt#1{\textit{#1}}
  \def\bt#1{#1}\def\bvol#1{\textbf{#1}} \def\vol#1{#1} \def\pg#1{#1}
  \def\publ#1{#1}\def\arxiv#1{#1}\def\org#1{#1}\def\st#1{\textit{#1}}

\bibitem[Bellamy(1982)]{Bellamy1982}
{\sc \au{Bellamy, N.~W.}} \yr{1982} Development of the {SEA Clam} wave energy
  converter.  \bt{In {\em Proceedings of the 2nd International Symposium on
  Wave Energy Utilization\/}},  \pg{pp. 175--190}. Trondheim, Norway.

\bibitem[Budal \& Falnes(1975)]{Budal1975}
{\sc \au{Budal, K.} \& \au{Falnes, J.}} \yr{1975}  \at{A resonant point
  absorber of ocean-wave power}.  \jt{Nature}  \bvol{256},  \pg{478--479}, with
  Corrigendum in Nature, vol.~257, p.~626, 1975.

\bibitem[Chaplin {\em et~al.\/}(2015{\natexlab{{\em a\/}}})Chaplin, Farley,
  Greaves, Hann, Kurniawan \& Cox]{Kurniawan2015b}
{\sc \au{Chaplin, J.}, \au{Farley, F.}, \au{Greaves, D.}, \au{Hann, M.},
  \au{Kurniawan, A.} \& \au{Cox, M.}} \yr{2015{\natexlab{{\em a\/}}}} Numerical
  and experimental investigation of wave energy devices with inflated bags.
  \bt{In {\em Proc. 11th Eur. Wave and Tidal Energy Conf.\/}}. Nantes, France.

\bibitem[Chaplin {\em et~al.\/}(2015{\natexlab{{\em b\/}}})Chaplin, Farley,
  Kurniawan, Greaves \& Hann]{Chaplin2015}
{\sc \au{Chaplin, J.~R.}, \au{Farley, F.}, \au{Kurniawan, A.}, \au{Greaves, D.}
  \& \au{Hann, M.}} \yr{2015{\natexlab{{\em b\/}}}} Forced heaving motion of a
  floating air-filled bag.  \bt{In {\em Proc. 30th Int. Workshop on Water Waves
  and Floating Bodies\/}}. Bristol, UK.

\bibitem[Chaplin {\em et~al.\/}(2012)Chaplin, Heller, Farley, Hearn \&
  Rainey]{Chaplin2012}
{\sc \au{Chaplin, J.~R.}, \au{Heller, V.}, \au{Farley, F. J.~M.}, \au{Hearn,
  G.~E.} \& \au{Rainey, R. C.~T.}} \yr{2012}  \at{Laboratory testing the
  {A}naconda}.  \jt{Phil. Trans. R. Soc. A}  \bvol{370}~(1959),  \pg{403--424}.

\bibitem[Crowley {\em et~al.\/}(2013)Crowley, Porter \& Evans]{Crowley2013}
{\sc \au{Crowley, S.}, \au{Porter, R.} \& \au{Evans, D.~V.}} \yr{2013}  \at{A
  submerged cylinder wave energy converter}.  \jt{Journal of Fluid Mechanics}
  \bvol{716}.

\bibitem[Evans(1976)]{Evans1976}
{\sc \au{Evans, D.~V.}} \yr{1976}  \at{A theory for wave power absorption by
  oscillating bodies}.  \jt{Journal of Fluid Mechanics}  \bvol{77}~(1),
  \pg{1--25}.

\bibitem[Evans \& Porter(2012)]{EvansPorter2012}
{\sc \au{Evans, D.~V.} \& \au{Porter, R.}} \yr{2012}  \at{Wave energy
  extraction by coupled resonant absorbers}.  \jt{Phil. Trans. R. Soc. A}
  \bvol{370}~(1959),  \pg{315--344}.

\bibitem[Falnes(2002)]{Falnes2002}
{\sc \au{Falnes, J.}} \yr{2002} {\em Ocean Waves and Oscillating Systems\/}.
  \publ{Cambridge, UK: Cambridge University Press}.

\bibitem[Farley(2011)]{Farley2011}
{\sc \au{Farley, F. J.~M.}} \yr{2011} The free floating clam - a new wave
  energy converter.  \bt{In {\em Proceedings of the 9th European Wave and Tidal
  Energy Conference\/}}. Southampton.

\bibitem[Farley(2012)]{Farley2012patent}
{\sc \au{Farley, F. J.~M.}} \yr{2012} Free floating bellows wave energy
  converter. {UK Patent GB2488185}.

\bibitem[Fenton(1978)]{Fenton1978}
{\sc \au{Fenton, J.~D.}} \yr{1978}  \at{Wave forces on vertical bodies of
  revolution}.  \jt{Journal of Fluid Mechanics}  \bvol{85},  \pg{241--255}.

\bibitem[French(1979)]{French1979}
{\sc \au{French, M.~J.}} \yr{1979} The search for low cost wave energy and the
  flexible bag device.  \bt{In {\em Proc. 1st Symp. Wave Energy
  Utilization\/}},  \pg{pp. 364--377}. Gothenburg, Sweden.

\bibitem[Gomes {\em et~al.\/}(2015{\natexlab{{\em a\/}}})Gomes, Henriques, Gato
  \& Falc\~ao]{Gomes2015b}
{\sc \au{Gomes, R. P.~F.}, \au{Henriques, J. C.~C.}, \au{Gato, L. M.~C.} \&
  \au{Falc\~ao, A. F.~O.}} \yr{2015{\natexlab{{\em a\/}}}}  \at{Testing of a
  small-scale model of a heaving floating {OWC} in a wave channel and
  comparison with numerical results}.  \bt{In {\em Renewable Energies
  Offshore\/} (ed. \ed{C.~Guedes Soares})},  \pg{pp. 445--454}.  \publ{CRC
  Press}.

\bibitem[Gomes {\em et~al.\/}(2015{\natexlab{{\em b\/}}})Gomes, Henriques, Gato
  \& Falc\~ao]{Gomes2015}
{\sc \au{Gomes, R. P.~F.}, \au{Henriques, J. C.~C.}, \au{Gato, L. M.~C.} \&
  \au{Falc\~ao, A. F.~O.}} \yr{2015{\natexlab{{\em b\/}}}} Wave channel tests
  of a slack-moored floating oscillating water column in regular waves.  \bt{In
  {\em Proceedings of the 11th European Wave and Tidal Energy Conference\/}}.
  Nantes, France.

\bibitem[Harrison(1970)]{Harrison1970}
{\sc \au{Harrison, H.~B.}} \yr{1970}  \at{The analysis and behaviour of
  inflatable membrane dams under static loading}.  \jt{Proceedings of the
  Institution of Civil Engineers}  \bvol{45}~(4),  \pg{661--676}.

\bibitem[Hulme(1982)]{Hulme1982}
{\sc \au{Hulme, A.}} \yr{1982}  \at{The wave forces acting on a floating
  hemisphere undergoing forced periodic oscillations}.  \jt{Journal of Fluid
  Mechanics}  \bvol{121},  \pg{443--463}.

\bibitem[Isaacson(1982)]{Isaacson1982}
{\sc \au{Isaacson, M. de St.~Q.}} \yr{1982}  \at{Fixed and floating
  axisymmetric structures in waves}.  \jt{Journal of the Waterway Port Coastal
  and Ocean Division}  \bvol{108},  \pg{180--199}.

\bibitem[Kurniawan {\em et~al.\/}(2014)Kurniawan, Greaves \&
  Chaplin]{Kurniawan2014a}
{\sc \au{Kurniawan, A.}, \au{Greaves, D.} \& \au{Chaplin, J.}} \yr{2014}
  \at{Wave energy devices with compressible volumes}.  \jt{Proceedings of the
  Royal Society of London A}  \bvol{470}~(2172).

\bibitem[Kurniawan {\em et~al.\/}(2016)Kurniawan, Greaves, Hann, Chaplin \&
  Farley]{Kurniawan2016}
{\sc \au{Kurniawan, A.}, \au{Greaves, D.}, \au{Hann, M.}, \au{Chaplin, J.~R.}
  \& \au{Farley, F.}} \yr{2016} Wave energy absorption by a floating air-filled
  bag.  \bt{In {\em Proc. 31st Int. Workshop on Water Waves and Floating
  Bodies\/}}. Plymouth, US.

\bibitem[Lopes \& Sarmento(2002)]{Lopes2002}
{\sc \au{Lopes, D. B.~S.} \& \au{Sarmento, A. J. N.~A.}} \yr{2002}
  \at{Hydrodynamic coefficients of a submerged pulsating sphere in finite
  depth}.  \jt{Ocean Engineering}  \bvol{29}~(11),  \pg{1391--1398}.

\bibitem[Newman(1962)]{Newman1962}
{\sc \au{Newman, J.~N.}} \yr{1962}  \at{The exciting forces on fixed bodies in
  waves}.  \jt{Journal of Ship Research}  \bvol{6}~(3),  \pg{10--17}.

\bibitem[Newman(1976)]{Newman1976}
{\sc \au{Newman, J.~N.}} \yr{1976} The interaction of stationary vessels with
  regular waves.  \bt{In {\em Eleventh Symposium on Naval Hydrodynamics\/}},
  \pg{pp. 491--501}. London.

\bibitem[Newman(1994)]{Newman1994}
{\sc \au{Newman, J.~N.}} \yr{1994}  \at{{Wave effects on deformable bodies}}.
  \jt{Applied Ocean Research}  \bvol{16},  \pg{47--59}.

\bibitem[Pagitz(2007)]{Pagitz2007}
{\sc \au{Pagitz, M.}} \yr{2007}  \at{The future of scientific ballooning}.
  \jt{Phil. Trans. R. Soc. A}  \bvol{365}~(1861),  \pg{3003--3017}.

\bibitem[Pagitz \& Pellegrino(2010)]{Pagitz2010}
{\sc \au{Pagitz, M.} \& \au{Pellegrino, S.}} \yr{2010}  \at{Maximally stable
  lobed balloons}.  \jt{Int. Journal of Solids and Structures}
  \bvol{47}~(11--12),  \pg{1496--1507}.

\bibitem[Parbery(1976)]{Parbery1976}
{\sc \au{Parbery, R.~D.}} \yr{1976}  \at{A continuous method of analysis for
  the inflatable dam.}  \jt{Proceedings of the Institution of Civil Engineers}
  \bvol{61}~(4),  \pg{725--736}.

\bibitem[Pimm {\em et~al.\/}(2014)Pimm, Garvey \& de~Jong]{Pimm2014}
{\sc \au{Pimm, A.~J.}, \au{Garvey, S.~D.} \& \au{de~Jong, M.}} \yr{2014}
  \at{Design and testing of energy bags for underwater compressed air energy
  storage}.  \jt{Energy}  \bvol{66},  \pg{496 -- 508}.

\bibitem[Taylor(1963)]{Taylor1919}
{\sc \au{Taylor, G.~I.}} \yr{1963}  \at{On the shapes of parachutes}.  \bt{In
  {\em The Scientific Papers of G. I. Taylor\/} (ed. \ed{G.~K. Batchelor})},
  \pg{pp. 26--37}.  \publ{Cambridge University Press}, (Original work published
  1919).

\bibitem[Todalshaug {\em et~al.\/}(2016)Todalshaug, {\'A}sgeirsson,
  Hj{\'a}lmarsson, Maillet, M{\"o}ller, Pires, Gu{\'e}rinel \&
  Lopes]{Todalshaug2016}
{\sc \au{Todalshaug, J.~H.}, \au{{\'A}sgeirsson, G.~S.}, \au{Hj{\'a}lmarsson,
  E.}, \au{Maillet, J.}, \au{M{\"o}ller, P.}, \au{Pires, P.}, \au{Gu{\'e}rinel,
  M.} \& \au{Lopes, M.}} \yr{2016}  \at{Tank testing of an inherently
  phase-controlled wave energy converter}.  \jt{International Journal of Marine
  Energy}  \bvol{15},  \pg{68--84}.

\bibitem[WAMIT(2015)]{WAMIT7.1}
{\sc \au{WAMIT}} \yr{2015} WAMIT, Inc., Chestnut Hill, MA, version 7.1.

\bibitem[Zhang {\em et~al.\/}(2014)Zhang, Yang \& Xiao]{Zhang2014}
{\sc \au{Zhang, X.}, \au{Yang, J.} \& \au{Xiao, L.}} \yr{2014}  \at{Numerical
  study of an oscillating wave energy converter with nonlinear snap-through
  power-take-off systems in regular waves}.  \jt{Journal of Ocean and Wind
  Energy}  \bvol{1}~(4),  \pg{225--230}.

\end{thebibliography}

\end{document}